\documentclass[useAMS,usenatbib]{mnras}	%Two columns
\usepackage[figuresright]{rotating}
\usepackage{pdflscape} 

\usepackage{graphics}

\usepackage{bigdelim}
\usepackage{bigstrut}

\usepackage{setspace}

\usepackage{natbib}
\usepackage{mathtools}
\usepackage{amsmath}
\usepackage{amssymb}
\usepackage[subnum]{cases}

\usepackage{times}
\usepackage{xspace}

\newcommand{\Msun}{\,{\rm M}_{\sun}}

\newcommand{\Gyr}{\,{\rm Gyr}}

\newcommand{\pc}{\,{\rm pc}}

\newcommand{\ZH}{\lbrack {\rm Z/H} \rbrack}

\newcommand{\Mcstar}{M_{\rm c,\ast}}

\newcommand{\Mgmc}{M_{\rm GMC}}

\newcommand{\emosaics}{{\sc E-MOSAICS}\xspace}
\newcommand{\mosaics}{{\sc MOSAICS}\xspace}
\newcommand{\eagle}{{\sc EAGLE}\xspace}

\title[The origin of the blue tilt]{The origin of the `blue tilt' of globular cluster populations in the E-MOSAICS simulations}

\author[Usher et al.]{Christopher Usher,$^{1}$\thanks{email: C.G.Usher@ljmu.ac.uk}
Joel Pfeffer,$^{1}$
Nate Bastian,$^{1}$
J. M. Diederik Kruijssen,$^{2}$ \newauthor
Robert A. Crain$^{1}$ and
Marta Reina-Campos$^{2}$
\\
  $^1$Astrophysics Research Institute, Liverpool John Moores University, 146 Brownlow Hill, Liverpool L3 5RF, UK \\
  $^2$Astronomisches Rechen-Institut, Zentrum f\"ur Astronomie der Universit\"at Heidelberg, M\"onchhofstra{\ss}e 12-14, D-69120 Heidelberg, Germany
}

\begin{document}

\maketitle

\begin{abstract}
The metal-poor sub-population of globular cluster (GC) systems exhibits a correlation between the GC average colour and luminosity, especially in those systems associated with massive elliptical galaxies.
More luminous (more massive) GCs are typically redder and hence more metal-rich.
This `blue tilt' is often interpreted as a mass-metallicity relation stemming from GC self-enrichment, whereby more massive GCs retain a greater fraction of the enriched gas ejected by their evolving stars, fostering the formation of more metal-rich secondary generations.
We examine the E-MOSAICS simulations of the formation and evolution of galaxies and their GC populations, and find that their GCs exhibit a colour-luminosity relation similar to that observed in local galaxies, without the need to invoke mass-dependent self-enrichment.
We find that the blue tilt is most appropriately interpreted as a dearth of massive, metal-poor GCs: the formation of massive GCs requires high interstellar gas surface densities, conditions that are most commonly fostered by the most massive, and hence most metal rich, galaxies, at the peak epoch of GC formation.
The blue tilt is therefore a consequence of the intimate coupling between the small-scale physics of GC formation and the evolving properties of interstellar gas hosted by hierarchically-assembling galaxies.
\end{abstract}


\section{Introduction}
\label{sec:introduction}

Virtually all galaxies with stellar masses greater than $10^{9}$ M$_{\sun}$ host populations of globular clusters (GCs) \citep[see reviews by ][]{2006ARA&A..44..193B, 2018RSPSA.47470616F}.
Mostly old \citep[ages $> 10$ Gyr, formation redshifts $z > 2$, e.g.][]{2005A&A...439..997P, 2005AJ....130.1315S, 2013ApJ...775..134V}, the number of these dense \citep[average half light radii of $\sim 3$ pc, e.g.][]{2010ApJ...715.1419M}, massive \citep[average masses of $\sim 2 \times 10^{5} \Msun$, e.g.][]{2007ApJS..171..101J} star clusters increases with galaxy mass \citep[and references therein]{2013ApJ...772...82H} with the Milky Way hosting $\sim 160$ GCs \citep{1996AJ....112.1487H, 2010arXiv1012.3224H} and the brightest cluster galaxy M87 hosting $\sim13$ 000 GCs \citep{2008ApJ...681..197P}.
Since the properties of GCs reflect both the conditions of their formation and of their survival to the present \citep[e.g.][]{2014CQGra..31x4006K}, GCs have long been used to study how galaxies form and evolve \citep[e.g.][]{1978ApJ...225..357S, 1996AJ....111.1529G, 2005ApJ...623..650K, 2008ApJ...681..197P, 2010MNRAS.404.1203F, 2014ApJ...796...52B, KPRCB18}.

The metallicities of extragalactic GCs have traditionally been studied using optical photometry.
Most \citep[e.g.][]{2001AJ....121.2974L, 2006ApJ...639...95P}, but not all \citep[e.g.][]{2017ApJ...835..101H}, massive galaxies show bimodal distributions of GC colours.
Most extragalactic GC studies of the last two decades have been analysed and interpreted through the lens of these two colour subpopulations.
This colour bimodality has traditionally been interpreted as a metallicity bimodality and has been used to motivate various two phase models of galaxy formation \citep[e.g.][]{1992ApJ...384...50A, 1997AJ....113.1652F, 1998ApJ...501..554C}.
This interpretation of the subpopulations as distinct entities is supported by their generally differing kinematics and spatial distributions \citep[e.g.][]{2013MNRAS.428..389P} and by the observation of clear GC colour bimodality and metallicity bimodality in some galaxies \citep[e.g. NGC 3115,][]{2012ApJ...759L..33B}.

The `blue tilt' is an observed optical colour-magnitude relation where the brightest blue (metal-poor) GCs are redder (more metal-rich) than the fainter blue GCs.
First reported by \citet{2006ApJ...636...90H}, \citet{2006AJ....132.1593S}, \citet{2006AJ....132.2333S} and \citet{2006ApJ...653..193M}, blue tilts are seen in most but not all massive galaxies (e.g. NGC 4472 \citealt{2006AJ....132.2333S} and UGC 10143 \citealt{2017ApJ...835..101H} lack blue tilts).
Hints of the blue tilt were reported earlier by \citet{1998AJ....116.2854O} and \citet{2003AJ....125.1908D}, who found evidence that the brightest GCs in NGC 1399 have a different (unimodal) colour distribution compared to the GC population as a whole.

The blue tilt is seen in galaxies across a wide range of environments, from massive galaxy clusters \citep[e.g.][]{2006ApJ...636...90H, 2014A&A...567A.105F, 2017ApJ...835..101H} to the field \citep[e.g.][]{2010MNRAS.401.1965H, 2014AJ....148...32J}, and with a wide range of morphologies, from ellipticals to spirals \citep{2009RAA.....9..993F, 2010MNRAS.403..429F}.
In observational studies, the blue tilt is quantified as the change in colour as a function of magnitude, so a `stronger' or `steeper' blue tilt corresponds to a greater change in colour with luminosity.

Stacking the GC systems of galaxies in the Virgo and the Fornax clusters, \citet{2006ApJ...653..193M, 2010ApJ...710.1672M} found evidence that the blue tilt is stronger in more massive galaxies and is weaker or non-existent in galaxies with stellar masses less than $\sim 10^{10}$ M$_{\sun}$.
However, as shown by a comparison of the GC systems of NGC 4472, NGC 4486 and NGC 4649 in the Virgo cluster, there is a diversity of blue tilts at fixed galaxy stellar mass \citep{2006AJ....132.2333S, 2006ApJ...653..193M, 2009AJ....138..758C}. 
In addition to differences in the strength of the blue tilt, the shape of the blue tilt varies galaxy-to-galaxy, with the colour-magnitude relation steepening at brighter GC luminosities in some galaxies, but not others \citep[e.g.][]{2009ApJ...699..254H, 2009ApJ...703...42P, 2009AJ....138..758C, 2010ApJ...710.1672M, 2017ApJ...835..101H}.
The blue tilt appears to be stronger in the centres of galaxies than in their outskirts in most \citep{2010ApJ...710.1672M, 2012MNRAS.420...37B, 2013MNRAS.436.1172U} but not all studies \citep{2009ApJ...699..254H, 2011MNRAS.413.2943F}.
The blue tilt is seen in a wide range of colours, including UV-optical colours (F275W $- V$ \citealt{2015ApJ...805..178B}), optical colours (e.g. $B - I$ \citealt{2009ApJ...699..254H}, $g - z$ \citealt{2006AJ....132.2333S}), and optical-near-infrared colours ($I - H$, \citealt{2012ApJ...746...88B}, $R - [3.6]$ \citealt{2008MNRAS.389.1150S}).

Whereas a colour-magnitude relationship is prevalent among blue GCs, evidence for such a relation in the red subpopulation is inconclusive, with a range of colour-magnitude relations observed \citep[e.g.][]{2009ApJ...699..254H, 2010ApJ...710.1672M, 2013MNRAS.436.1172U}; any `red tilt' is weaker than the blue tilt.
Caution is warranted for these claims as the process of measuring the peaks of the colour distribution can introduce (anti-)correlations between the colour-magnitude relationship of each subpopulation \citep{2006ApJ...653..193M, 2009ApJ...699..254H}.

The blue tilt is usually interpreted as a mass-metallicity relation since colour traces metallicity and luminosity traces mass.
This interpretation is supported by spectroscopic studies \citep{2010AJ....139.1566F, 2012MNRAS.426.1475U}, with \citet{2015MNRAS.446..369U} finding a metallicity-luminosity relation consistent with the colour-magnitude relationship of the same GCs.
This phenomenon can be seen as a lack of the most metal-poor GCs at the highest masses.
As will be shown in this paper, this interpretation differs fundamentally from one in which the blue tilt reflects a shift in colour or metallicity of the most massive GCs.
Since the blue tilt is a second order effect, a large sample of GCs (several hundred) is required to detect its presence.
Thus, the GC system of the Milky Way is too small \citep[$\sim 160$ GCs,][]{1996AJ....112.1487H, 2010arXiv1012.3224H} to observe the presence or absence of a blue tilt.

The translation of observed colour-magnitude relations into mass-metallicity relations is complicated by a few effects.
Firstly, the relationship between colour and metallicity is non-linear for old stellar populations \citep[e.g.][]{2006ApJ...639...95P, 2006Sci...311.1129Y, 2009ApJ...699..486C, 2012MNRAS.426.1475U}.
This means that a different luminosity-metallicity relation can be found depending on whether a colour-magnitude relation is first fit before being converted into a luminosity-metallicity relation, or if the colours are first converted into metallicities before fitting the luminosity-metallicity relation.
Secondly, there is now observational evidence \citep{2012MNRAS.426.1475U, 2015MNRAS.446..369U, 2016ApJ...829L...5P, 2017ApJ...844..104P} that the GC colour-metallicity relation varies between galaxies, with some galaxies showing nearly linear colour-metallicity relations and others display highly non-linear relations.
This is to be expected, because GC colours depend on their age and chemistry, and galaxies exhibit a range of formation histories and thus would have different age-metallicity and chemistry-metallicity relationships in their GC populations.
Thus, two galaxies with similar colour-magnitude relations could have different metallicity-mass relations.
Differences in the colour-metallicity relationship within galaxies would further complicate the interpretation of the blue tilt, although \citet{2015MNRAS.446..369U} found no evidence that the colour-metallicity relation varies with GC luminosity.

Thirdly, the conversion of absolute magnitudes into masses is also complicated by the uncertainty in mass-to-light ratios ($M/L$).
The effect of metallicity on ($M/L$) is debated as stellar population models \citep[e.g.][]{2010ApJ...712..833C} predict that ($M/L$) increases with metallicity, while dynamical studies of GCs in the Milky Way \citep{2005ApJS..161..304M, 2017MNRAS.464.2174B} and M31 \citep{2011AJ....142....8S} show that the $V$-band ($M/L$) is largely independent of metallicity.
This difference may be reconciled by dynamical effects such as disruption or mass segregation \citep[e.g.][]{2008A&A...486L..21K, 2015MNRAS.448L..94S}.
Another concern is whether the ($M/L$) varies with cluster mass.
Due to mass segregation, GCs should preferentially lose their low mass stars, causing their ($M/L$) to decrease \citep{2003MNRAS.340..227B, 2009A&A...500..785K}.
This effect is stronger for lower mass GCs and for GCs in denser environments.
Observationally, some studies find evidence for a decrease in ($M/L$) at low masses \citep[e.g.][]{2011AJ....142....8S, 2015AJ....149...53K} while others do not \citep[e.g.][]{2016ApJ...833....8G, 2018MNRAS.tmp.1027B}.
We note that the scatter in the ($M/L$) of Milky Way GCs at fixed mass and metallicity is larger than observational uncertainties \citep
[e.g.][]{2018MNRAS.tmp.1027B}.

Although various models, including contamination by stripped galactic nuclei \citep[e.g.][]{2006ApJ...636...90H, 2006ApJ...653..193M}, a bottom heavy initial mass function \citep{2014ApJ...780...43G}, and self-enrichment \citep[e.g.][]{2008AJ....136.1828S, 2009ApJ...695.1082B}, have been proposed to explain the blue tilt phenomenon, each of these models has serious limitations in light of recent advances in observations of old GCs \citep[e.g.][]{2009A&A...508..695C, 2012ApJS..200....4F, 2014ApJ...796...52B} and of the formation of high mass stellar clusters in the Local Universe \citep[e.g][]{2010ARA&A..48..431P, 2014MNRAS.441.2754C,2014prpl.conf..291L}.
The recent advent of cosmological, hydrodynamical simulations of galaxies that also incorporate self-consistent treatments of the formation and evolution of their star cluster populations \citep{2018MNRAS.475.4309P, K18} therefore makes this a judicious time to re-examine the origin of the blue tilt.

This paper is organised as follows.
First, we discuss the current theories for the origin of the blue tilt and their limitations in Section~\ref{sec:old_theories}.
In Section~\ref{sec:emosaics}, we will then look for the presence of the blue tilt in the \emosaics simulations of the formation and evolution of globular cluster systems and investigate its physical origin.
In Section \ref{sec:colour_mag}, we study the behaviour of the blue tilt in \emosaics using the same colour-magnitude relation analysis used in observational studies, before giving our conclusions in Section~\ref{sec:conclusion}.

\section{Existing theories for the blue tilt}
\label{sec:old_theories}

Several scenarios have been proposed to explain the origin of the blue tilt.
The most common explanation for the blue tilt is that it is the result of self-enrichment \citep{2008AJ....136.1828S, 2009ApJ...695.1082B}.
In this scenario, massive star clusters are able to retain the metals (including heavier elements such as Ca and Fe) synthesised by their massive stars and to form subsequent generation(s) of enriched stars, with higher mass clusters being able to retain a larger fraction of the newly created metals and self-enrich to a greater extent.  
This model for the blue tilt can be seen as an extension of the self-enrichment models for multiple populations of chemically distinct stars within GCs \citep[e.g.][]{2007A&A...464.1029D, 2008MNRAS.391..825D} to higher GCs masses \citep[e.g.][]{2011A&A...533A.120V}. 
In a self-enrichment scenario, the absence of a red tilt is due to the smaller relative change in metallicity at high metallicity for the addition of the same mass of metals.
Since a minimum binding energy (and hence mass, given the weak mass-radius relation of GCs) is required to retain supernova ejecta, the self-enrichment model explains why the blue tilt weakens or disappears below a certain minimum GC mass.
The \citet{2009ApJ...695.1082B} model allows for variation in the blue tilt by changing the star formation efficiency, the density profiles of the protocluster gas clouds or the initial star cluster radius.
Some studies \citep[e.g.][]{2010ApJ...710.1672M, 2014A&A...567A.105F} have used the \citet{2009ApJ...695.1082B} model and observations to attempt to constrain the conditions of GC formation in different environments.

The self-enrichment model requires GC formation to last long enough for the bulk of the metals produced by the GC's high mass stars to be incorporated into the low mass stars that are luminous today.
This is a major issue for the model, because young massive clusters (YMCs) with masses approaching $10^{8}$ M$_{\sun}$ \citep{2004A&A...416..467M} show no evidence for extended star formation histories \citep{2013MNRAS.436.2852B, 2014MNRAS.441.2754C, 2016MNRAS.457..809C} or gas or dust \citep{2014MNRAS.443.3594B, 2015MNRAS.448L..62L}. 
Furthermore, YMCs are gas free on timescales ($< 4$ Myr) shorter than the lifetimes of the progenitors of core collapse supernovae \citep{2014MNRAS.445..378B, 2014prpl.conf..291L, 2015MNRAS.449.1106H}.
Besides the lack of observational evidence for extended star formation in YMCs, simulations of star cluster formation also suggest timescales of $\lesssim 5$ Myr \citep{2012MNRAS.420.1457H, 2017ApJ...834...69L}. 
Another issue for the self-enrichment scenario is that to explain the variation in the blue tilt between galaxies, the initial structure of GCs must vary between galaxies.

The self-enrichment scenario also predicts large metallicity spreads within GCs, with massive GCs having metallicity spreads of at least 0.5 dex.
Beyond $\omega$ Cen \citep[e.g.][]{2010ApJ...722.1373J}, M 54 \citep[e.g.][]{2010A&A...520A..95C} and Terzan 5 \citep[e.g.][]{2014ApJ...795...22M}, Milky Way GCs show either small ($\Delta$ [Fe/H] $< 0.2$) or non-existent metallicity spreads \citep[see e.g.][]{2017arXiv171201286B}.
We note that $\omega$ Cen is likely the stripped nucleus of a galaxy accreted onto the Milky Way \citep[e.g][]{1999Natur.402...55L, 2000A&A...362..895H}, M54 is the nucleus of Sagittarius dwarf galaxy currently undergoing such a process \citep[e.g][]{1997AJ....113..634I, 2008AJ....136.1147B, 2010ApJ...714L...7C} and Terzan 5 is metal-rich \citep{2014ApJ...795...22M}.
We note that many of the claims of small metallicity spreads in Milky Way GCs are likely due to spurious \ion{Fe}{i} based measurements \citep{2015ApJ...809..128M}.

An early alternative suggestion was that contamination by the stripped nuclei of galaxies could produce the blue tilt \citep{2006ApJ...636...90H, 2006ApJ...653..193M}.
Like galaxies, galaxy nuclei follow a mass-metallicity relation, with more massive galactic nuclei being more metal-rich \citep[e.g.][]{2011MNRAS.413.1764P}.
Since the mass of the stripped nucleus depends on the initial nucleus mass \citep{2013MNRAS.433.1997P}, this would produce a mass-metallicity relationship for the stripped nuclei.
However, in the $10^{6}$ to $10^{7}$ M$_{\sun}$ mass range where the blue tilt is clearly observed, there are not enough stripped nuclei relative to GCs to explain the blue tilt \citep{2012A&A...537A...3M, 2014MNRAS.444.3670P, 2016MNRAS.458.2492P}.
Additionally, a stripped nuclei origin for the blue tilt cannot explain the lack of the brightest, bluest GCs \citep{2006ApJ...653..193M}.

Since lower mass GCs lose relatively more of their lowest mass stars, old low-mass GCs should appear bluer than than high mass GCs of the same age and metallicity.
While this effect is subtle for a \citet{2001MNRAS.322..231K} or \citet{2003PASP..115..763C} Milky Way-like IMF, it is significant for a bottom heavy IMF such as those claimed in the centres of some massive galaxies \citep[$\gtrsim 10^{11} \Msun$, e.g.][]{2010Natur.468..940V, 2012ApJ...760...71C, 2012Natur.484..485C}, potentially explaining the origin of the blue tilt \citep{2014ApJ...780...43G}.
However, stacked spectra of massive GCs do not show the same evidence for bottom-heavy IMFs as massive early-type galaxies \citep{2015MNRAS.446..369U}, nor do metal-rich GCs in M31 show evidence for bottom-heavy IMFs \citep{2017ApJ...850L..14V} and the blue tilt is observed in lower-mass galaxies ($< 5 \times 10^{10} \Msun$), which do not show evidence for a bottom heavy IMF \citep[e.g.][]{2012ApJ...760...71C, 2012Natur.484..485C, 2013MNRAS.433.3017L}.
While the effects of dynamical evolution on ($M/L$) should be considered in blue tilt studies, GCs are highly unlikely to have had extreme enough IMFs to explain the blue tilt through the mechanism proposed by \citet{2014ApJ...780...43G}.

\citet{2006ApJ...653..193M} attempted to explain the blue tilt by exploring whether the GC populations of large galaxies can be built up through merging of lower-mass galaxies since the mean colour and mean luminosity of GCs increase with galaxy mass \citep[e.g.][]{2007ApJS..171..101J, 2010ApJ...717..603V, 2006AJ....132.2333S, 2006ApJ...639...95P}.
However, the mean colour of blue GCs only varies weakly with galaxy luminosity \citep[e.g.][]{2006AJ....132.2333S, 2006ApJ...639...95P, 2011ApJ...728..116L} so, using the observed GC scaling relations, \citet{2006ApJ...653..193M} were unable to produce a blue tilt as strong as observed in the highest mass galaxies.

\citet{2006ApJ...636...90H} suggested a model where the blue tilt is due to a connection between GC mass and the mass of the molecular cloud the GC formed from.  
If the maximum molecular cloud mass increases with galaxy mass, the galaxy mass-metallicity relation \citep[e.g.][]{1979A&A....80..155L, 2005MNRAS.362...41G, 2013ApJ...779..102K, 2017ApJ...847...18Z} should produce a blue tilt.
This model has not been explored in a quantitative sense until now.
In the recent semi-analytic models of GC system formation of \citet{2018arXiv180103515C}, the maximum GC mass is calculated as a fraction of the molecular gas of the GC's host galaxy when the GC forms.
Since both the molecular gas mass and the metallicity increases with galaxy stellar mass in the \citet{2018arXiv180103515C} model, a blue tilt is produced.
Their resulting mass-metallicity relationship ($Z \propto M^{0.23 \pm 0.01}$) for their blue GCs is consistent with their analysis of GCs in the Virgo cluster and with some \citep[e.g.][]{2009AJ....138..758C, 2017ApJ...835..101H}, but not all \citep{2010ApJ...710.1672M}, observational studies.

\section{Blue tilts in E-MOSAICS}
\label{sec:emosaics}

In the this section we investigate the origin of the blue tilt with cosmological simulations of galaxy formation that include a physically-motivated, subgrid model for cluster formation and disruption.
By including or excluding physical processes in cluster formation, we can pinpoint the physics driving the blue tilt and determine if additional processes such as self-enrichment are needed.

\subsection{The E-MOSAICS models}
\label{sec:emosaics_model}

The MOdelling Star cluster population Assembly In Cosmological Simulations within \eagle \citep[\emosaics,][]{2018MNRAS.475.4309P, K18} project is a suite of cosmological, hydrodynamical simulations of galaxy formation in the $\Lambda$ cold dark matter cosmogony that couple the \mosaics model for star cluster formation and evolution \citep{2011MNRAS.414.1339K, 2018MNRAS.475.4309P} to the Evolution and Assembly of GaLaxies and their Environments \citep[\eagle,][]{2015MNRAS.446..521S, 2015MNRAS.450.1937C} simulations of galaxy formation and evolution. 
The simulations are run with a significantly modified version of the the $N$-body/smoothed particle hydrodynamics code \textsc{gadget3} \citep[last described by][]{2005MNRAS.364.1105S}.
The \eagle simulations are successful in reproducing a broad range of galaxy properties \citep[e.g.][]{2015MNRAS.450.4486F, 2017MNRAS.465..722F, 2015MNRAS.452.2879T, 2015MNRAS.452.3815L, 2017MNRAS.464.4204C, 2015MNRAS.452.2034R, 2017MNRAS.471..690T}.

The \emosaics simulations adopt baryonic particles with masses of $2.25 \times 10^5 \Msun$, thus resolving galaxies of $M_\star = 10^{10}\Msun$ with $\approx 40,000$ particles, and adopt the \eagle `Recal' model, which yields a better agreement with the observed galaxy population at this mass resolution than the \eagle reference model \citep{2015MNRAS.446..521S}.
Additionally, and particularly relevant for this work, the Recal model is in good agreement with the mass-metallicity relation of galaxies with $M_\ast \gtrsim 10^{8.5} \Msun$.

Galaxies are identified in the simulations in the same manner as in the \eagle simulations \citep[for details see][]{2015MNRAS.446..521S}.
Dark matter structures (FOF groups) are first identified using the friends-of-friends (FOF) algorithm using a linking length of 0.2 times the mean interparticle separation \citep{1985ApJ...292..371D}.
Bound galaxies (subhaloes) are then identified within the FOF groups using the \textsc{subfind} algorithm \citep{2001MNRAS.328..726S, 2009MNRAS.399..497D}.

The \mosaics model \citep{2011MNRAS.414.1339K, 2018MNRAS.475.4309P} adopts a subgrid approach, in which a fraction of a newly formed stellar particle is assumed to be formed in bound star clusters. The star clusters are `attached' to the stellar particles in the simulation and the formation properties of the star clusters are determined by resolved quantities in the hydrodynamical simulation. 
One of the benefits of this approach is that the subgrid models used by \eagle do not require recalibrating.
The subgrid prescriptions in \mosaics are based on models which reproduce direct $N$-body simulations of star cluster mass-loss and observed properties of young star clusters, such as the age distributions, mass distributions, spatial distributions, and kinematics \citep[e.g.][]{2011MNRAS.414.1339K, 2012MNRAS.421.1927K, 2017MNRAS.470.1421M, 2018ASSL..424...91A}. 
For further discussion, we refer the reader to \citet{2018MNRAS.475.4309P}.

Briefly, when a star particle forms, a subgrid population of stars is formed with a mass equal to the local cluster formation efficiency (CFE) $\Gamma$ \citep{2008MNRAS.390..759B} times the star particle mass.
The CFE is determined from the \citet{2012MNRAS.426.3008K} model according to the natal gas properties (gas density, velocity dispersion and sound speed).
The star cluster populations form with a \citet{1976ApJ...203..297S} mass function with a power-law slope $-2$ and an environmentally dependent exponential truncation mass. 
The truncation mass $\Mcstar = \epsilon \Gamma \Mgmc$ \citep{2014CQGra..31x4006K} depends on the star formation efficiency for an entire molecular cloud (assumed to be $\epsilon=0.1$), the CFE ($\Gamma$) and the mass of the molecular cloud $\Mgmc$.
As the \eagle simulations do not model the cold, dense phase of the interstellar medium, \emosaics adopts the model of \citet{2017MNRAS.469.1282R} which relates the maximum molecular cloud mass to the local \citet{1964ApJ...139.1217T} mass (which depends on the local dynamics, through the epicyclic frequency $\kappa$, and local gas properties) and decreases the mass scale further due to the effects of stellar feedback.
Cluster masses are stochastically sampled from the cluster mass function between masses of $100$ and $10^8 \Msun$, with the number of clusters determined by the mass of the stellar particle and the CFE.
Only clusters with masses greater than $5 \times 10^3 \Msun$ are evolved to reduce memory requirements.
All star clusters are assumed to have a constant half-mass radii of $4 \pc$.

Star clusters in the simulations lose mass through several channels, namely stellar evolution, two-body relaxation, tidal shocks and dynamical friction \citep[for details see][]{2011MNRAS.414.1339K, 2018MNRAS.475.4309P}.
Stellar evolutionary mass loss for clusters is proportional to that of the host stellar particle calculated in the \eagle model \citep[see][]{2015MNRAS.446..521S}. 
The mass loss from two-body relaxation and tidal shocks is calculated according to the local tidal field through the tidal tensor at the location of the star particle.
Finally, the removal of star clusters due to dynamical friction is treated in post-processing and applied at every snapshot (of which there are a total of 29 between $z=20$ and $z=0$).

Through their analysis of the \emosaics simulations, \citet{2018MNRAS.475.4309P} showed that GCs in Milky Way-like galaxies are consistent with being the surviving clusters of YMC-like formation physics acting in galaxies over cosmic time. 
They found that the YMC-based cluster formation reproduces the high mass end of the GC mass function and the maximum GC mass as a function of galactocentric radius in Milky Way-like galaxies, where the high mass end of the GC mass function is shaped by the exponential truncation mass $\Mcstar$ and dynamical friction.
However they also found that the simulations overpredicted the number of low mass GCs ($\lesssim 10^5 \Msun$), which was attributed to insufficient cluster disruption due to the lack of an explicitly modelled cold interstellar gas phase.
This shortcoming does not unduly affect the use of the simulations to study the blue tilt since we focus on the massive ($> 10^5 \Msun$) GC population.

\subsection{The blue tilt in E-MOSAICS}

We use the set of 25 zoom-in simulations of $L^\star$ ($M_\rmn{vir}=7\times10^{11}$--$3\times10^{12}~\Msun$) galaxies from \citet{2018MNRAS.475.4309P} and \citet{K18}, which represent an unbiased, volume-limited sample from the Recal-L025N0752 simulation presented in \citet{2015MNRAS.446..521S}.
From this set of simulations we select GCs in any galaxies that reside in FOF groups with less than 1 per cent of their mass comprised by low resolution collisionless particles (used to model the large-scale environment surrounding the high-resolution, `zoomed-in' region of the simulation).
We select all GCs with masses greater than $10^{5}$ M$_{\sun}$ and ages older than 8 Gyr (a redshift of $z \sim 1$).
This mass cut is motivated by the inefficient disruption of lower mass GCs in \emosaics, the fact that extragalactic surveys are generally limited to observing clusters with masses greater than a few times $10^5~\Msun$, and the fact that the blue tilt is found among GCs with masses $M>10^5~\Msun$.
Since Population III stars are not modelled in \eagle and galaxies with this low metallicity are usually poorly resolved, we exclude clusters more metal-poor than [Z/H] $= -3$.
This choice is supported by the lack of GCs more metal poor than [Z/H] $= -3$ in the Local Group \citep[e.g.][]{1996AJ....112.1487H, 2010arXiv1012.3224H, 2011AJ....141...61C, 2018MNRAS.tmp..821S}.
We only select GCs at galactocentric radii greater than 1 kpc.
Observational studies have trouble observing GCs at small galactocentric radii due to the high background surface brightness at the centres of massive galaxies, while GCs in \emosaics are overabundant at such radii due to insufficient disruption in the simulations \citep{2018MNRAS.475.4309P}.

These selection criteria result in a total of 10553 GCs.
Of these, 73 are associated with the 69 galaxies less massive than $10^{7}$ M$_{\sun}$, 691 with the 147 galaxies between $10^{7}$ and $10^{10}$ M$_{\sun}$, 5742 with 29 galaxies with stellar masses between $10^{10}$ M$_{\sun}$ and $3 \times 10^{10} \Msun$, and 4047 GCs with 8 galaxies with stellar mass above $3 \times 10^{10} \Msun$.
We note that a stellar mass of $10^{7}$ M$_{\sun}$ corresponds to $\sim 40$ star particles and as such the star formation histories of galaxies at this low mass are not well-sampled by our simulations.

To compare the predictions of \emosaics with observations, we transform the model masses and metallicities into colours and absolute magnitudes.
We use the empirical colour-metallicity relation of \citet{2012MNRAS.426.1475U} to transform the metallicity [Z/H] into $(g - z)$ and a constant mass-to-light ratio of 2 M$_{\sun}/$L$_{z,\sun}$ to convert mass into $M_{z}$.
We use the total metallicity [Z/H] rather than the iron abundance [Fe/H] since colours such as $(g - z)$ more closely trace the overall metallicity \citep[e.g.][]{2007MNRAS.382..498C, 2015MNRAS.449.1177V}.
In Appendix \ref{sec:various_cmrs} we show that using [Fe/H] rather than [Z/H] gives qualitatively similar results.
The choice of the Sloan Digital Sky Survey $g$ and $z$ bands \citep{1996AJ....111.1748F} is motivated by their similarity with the HST ACS F475W and F850LP filters commonly used in studies of the blue tilt \citep[e.g.][]{2006ApJ...653..193M, 2010ApJ...710.1672M} and with the HST ACS F435W and F814W filters used in other important blue tilt studies \citep[i.e.][]{2009ApJ...699..254H}.
In Appendix \ref{sec:various_cmrs} we find qualitatively similar blue tilts in $(g - z)$, $(g - i)$ and $(B - I)$.
The use of a ($M/L$) independent of metallicity is a compromise between GC observations which show a near-infrared ($M/L$) ratio declining with increasing metallicity \citep[e.g.][]{2011AJ....142....8S, 2017MNRAS.464.2174B} and stellar population models that predict a ($M/L$) ratio increasing with metallicity \citep[e.g.][]{2009ApJ...699..486C, 2012MNRAS.424..157V}.
In Appendix \ref{sec:various_cmrs} we consider the effects of other transformations from mass and metallicity into photometry and find little qualitative effect on our results.

\begin{figure*}
\begin{center}
\includegraphics[width=504pt]{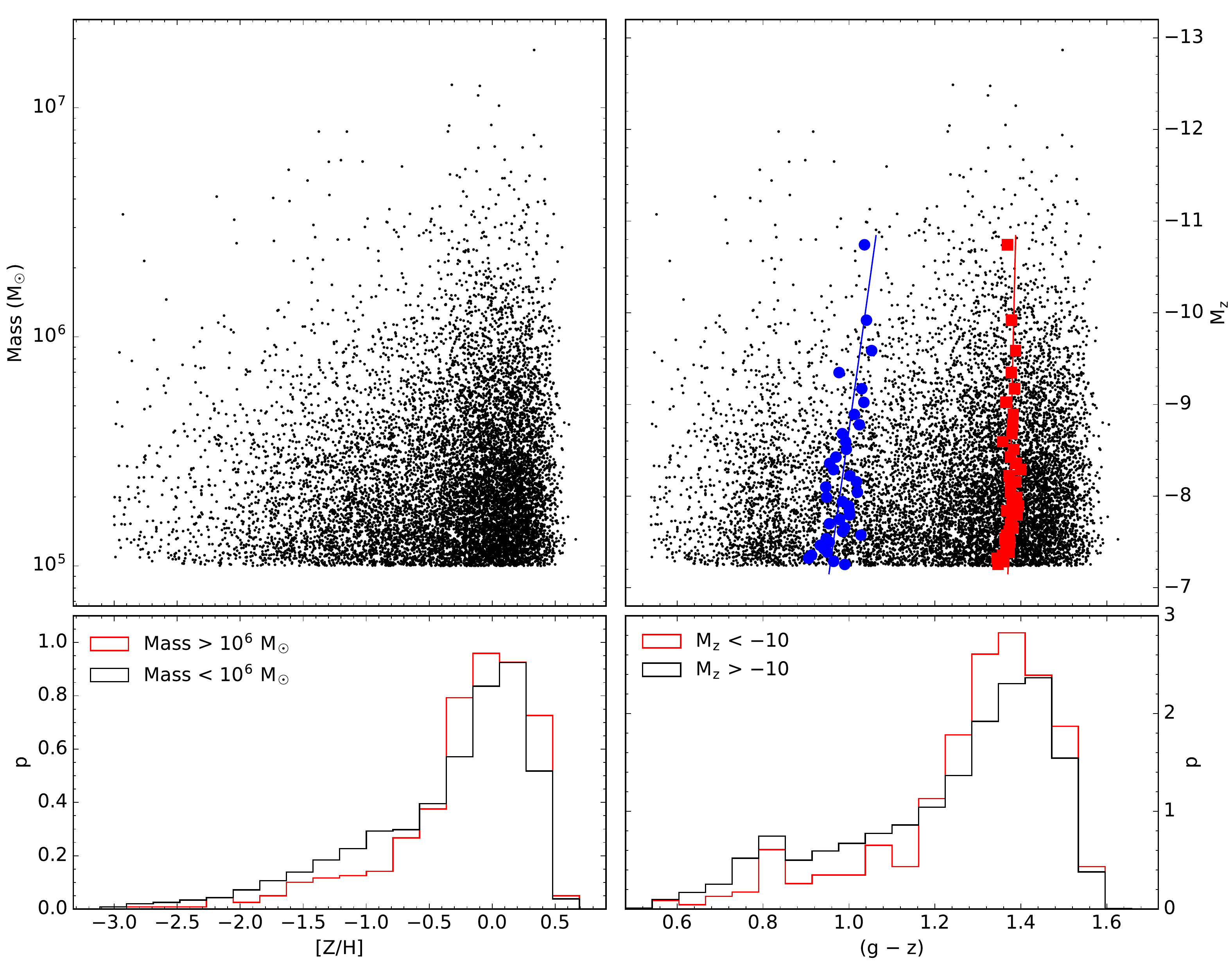}
\caption{The blue tilt in \emosaics.
\emph{Top left}: Mass-metallicity diagram for our sample of GCs in \emosaics.
Each point corresponds to a single GC.
The limits of the mass-metallicity space are the same as the transformed limits of the colour-magnitude diagram to the right.
\emph{Bottom left}: Metallicity distributions for GCs less massive than $10^{6}$ M$_{\sun}$ (black) and more massive than $10^{6}$ M$_{\sun}$ (red).
\emph{Top right}: Colour-magnitude diagram.
The blue circles and red squares show the means of two Gaussians fit to the colour distribution in absolute magnitude bins each with equal numbers of GCs.
The blue and red lines are least squares fits to these mean colours as a function of absolute magnitude.
The absolute magnitudes were calculated assuming a constant mass-to-light ratio of 2 M$_{\sun}/$L$_{z,\sun}$ and the colours were calculated using the empirical colour-metallicity relation of \citet{2012MNRAS.426.1475U}.
The pileup of points at $(g - z) = 0.84$ is due to the change of slope of the  \citet{2012MNRAS.426.1475U} relation at this colour, does not have a significant effect on the fitted blue tilt and is included in our fits to the colour distributions.
\emph{Bottom right}: Colour distribution for GCs brighter than $M_{z} = -10$ (black) and fainter than $M_{z} = -10$ (red).
There is a clear lack of metal-poor (blue) GCs at high masses (bright magnitudes).
The blue colour-magnitude relation has a slope of $-0.0279_{-0.0054}^{+0.0069}$. By contrast, the red colour-magnitude relation is very weak, with a slope of $-0.0049_{-0.0024}^{+0.0020}$.}
\label{fig:all_blue_tilt}
\end{center}
\end{figure*}

In the upper left panel of Figure \ref{fig:all_blue_tilt}, we show the mass-metallicity distribution for our GCs selected from \emosaics while in the upper right panel we show a colour-magnitude diagram for the same GCs.
There is a lack of massive, metal poor GCs or equivalently a lack of bright, blue GCs -- a blue tilt.
This difference in metallicity distribution (or colour distribution) between more and less massive GCs (or between brighter and fainter GCs) is illustrated by the histograms below the mass-metallicity plot in Figure \ref{fig:all_blue_tilt} (and the colour distributions below the colour-magnitude diagram in the same figure).
A Kolmogorov-Smirnov test gives a probability of $7 \times 10^{-8}$ that the metallicity distribution of GCs more massive than $10^{6}$ M$_{\sun}$ is the same as the metallicity distribution of GCs less massive than $10^{6}$ M$_{\sun}$.
Likewise, the colour distribution of GCs brighter than $M_{z} = -10$ shows a Kolmogorov-Smirnov probability of $3 \times 10^{-6}$ that it is drawn from the same colour distribution as GCs fainter than $M_{z} = -10$.

\subsection{The origin of the blue tilt in E-MOSAICS}
\label{sec:origin_blue_tilt}

In \emosaics, the maximum cluster mass is controlled by the exponential truncation mass of the initial cluster mass function, which is considered to be proportional to the maximum mass that can collapse into a molecular cloud, $M_{\rm c,*} = \varepsilon\Gamma M_{\rm GMC, max}$ \citep{2014CQGra..31x4006K}.
We determine the maximum cloud mass scale using the model of \citet{2017MNRAS.469.1282R}, which considers the balance between two physical mechanisms.
Centrifugal forces set the maximum mass that could collapse (i.e.~Toomre mass) in the two-dimensional free-fall time-scale of the galactic mid-plane.
However, due to the hierarchical structure of the interstellar medium star formation is initiated during the collapse, such that it can be halted by stellar feedback before a Toomre mass has collapsed.
In environments of low gas surface density with weak centrifugal forces stellar feedback is efficient at stopping the collapse, explaining why clouds are less massive in the local Universe than at high-redshift.
It reproduces observations showing that the maximum young star cluster mass varies galaxy-to-galaxy \citep[e.g.][Pfeffer et al. in prep.]{2006A&A...450..129G, 2009A&A...494..539L, 2012MNRAS.419.2606B} and correlates with star formation rate surface density \citep{2015MNRAS.452..246A, 2017ApJ...839...78J}.

Because the masses of the most massive GCs in \emosaics are regulated by the above model for the initial maximum cluster mass, we propose a new explanation for the origin of the blue tilt, in which the lack of massive, metal-poor GCs is caused by a physical upper limit in cluster formation ($\Mcstar$) that increases with increasing metallicity.
As a massive galaxy grows in mass and enriches in metallicity, the star forming gas reaches higher surface densities (pressures) due to the deeper potential of the galaxy, resulting in more massive clusters at higher metallicities.
However, low-mass galaxies with shallower potentials do not reach such high gas surface densities, resulting in maximum cluster masses that are approximately constant with metallicity.

This explanation for the blue tilt is supported by observations that the GC mass function is galaxy mass dependent, with higher mass galaxies possessing GC populations that extend to higher GC masses compared to low mass galaxies \citep[e.g][]{2007ApJS..171..101J, 2010ApJ...717..603V}, as well as by observations that show that the mean metallicities of GC systems increase with galaxy mass \citep[e.g.][]{2006ApJ...639...95P}.

\begin{figure*}
\begin{center}
\includegraphics[width=504pt]{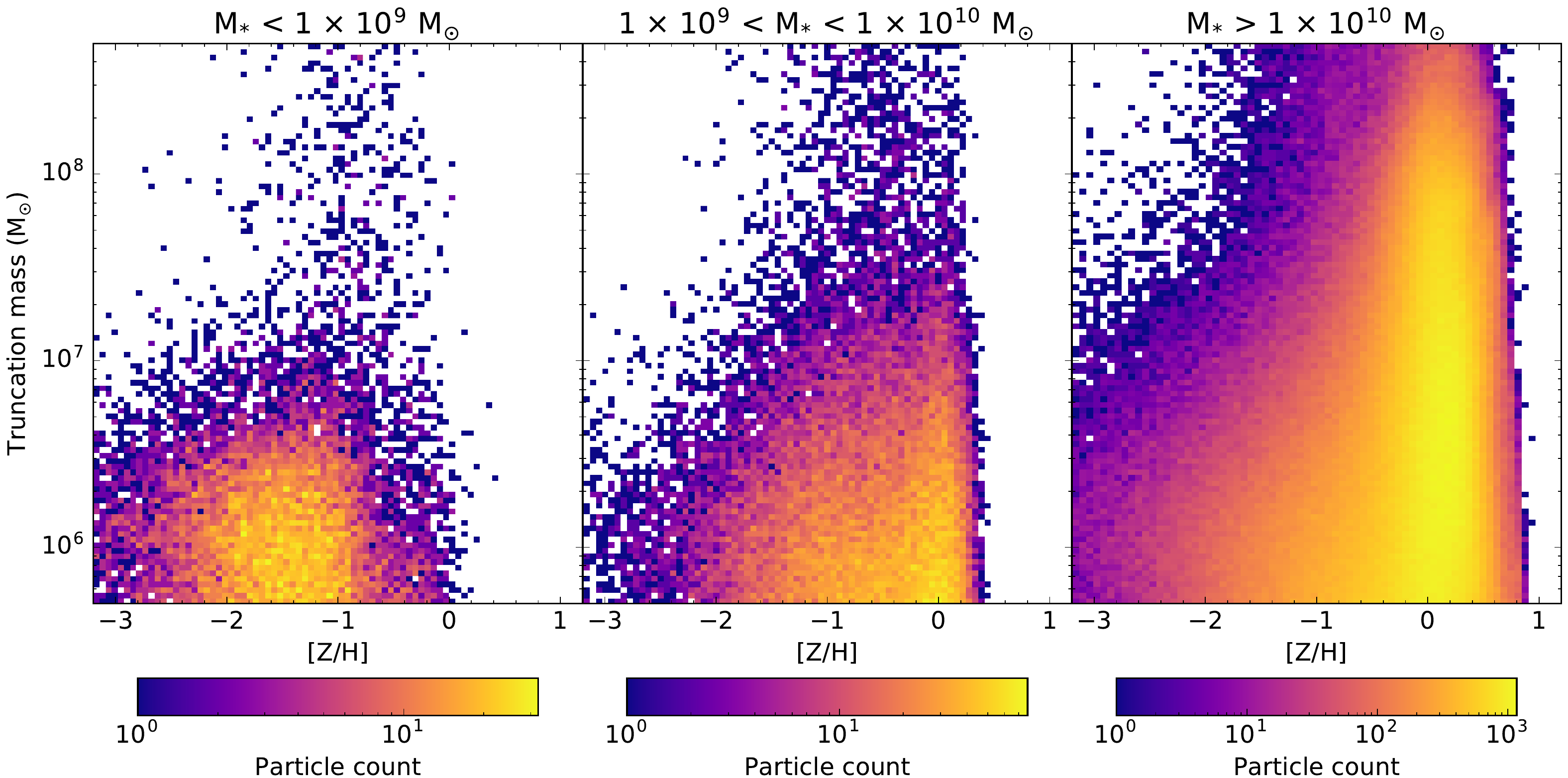}
\caption{Exponential truncation mass $\Mcstar$ as a function of metallicity for all star particles in galaxies with stellar mass less than $10^{9}$ M$_{\sun}$ (left panel), between $10^{9}$ M$_{\sun}$ and $10^{10}$ M$_{\sun}$ (middle) and more massive than $10^{10}$ M$_{\sun}$ (right).
The histogram bins are coloured such that the colours change from purple to yellow logarithmically with the number of particles in the bins.
As for the present day GC masses (Figure \ref{fig:all_blue_tilt}, there is a clear increase in maximum mass with increasing metallicity.
The small number of particles with high truncation masses in the low galaxy mass bins are mostly due numerical limitations in the calculation of the \citet{1964ApJ...139.1217T} mass.
More massive galaxies reach higher exponential truncation masses and higher metallicities.
As few GCs are formed more massive than the truncation mass, the conditions to form the most massive GCs are only present at high metallicities and in more massive galaxies.}
\label{fig:mcstar}
\end{center}
\end{figure*}

In Figure \ref{fig:mcstar}, we show the outcome of this model in the \emosaics simulations.
The figure shows the truncation mass as a function of metallicity for star particles in \emosaics galaxies in three galaxy stellar mass bins: $M_\ast < 10^9 \Msun$, $10^9 < M_\ast/\Msun < 10^{10}$ and $M_\ast > 10^{10} \Msun$.
We note that, since GCs in the simulations typically lose $\ga50$ per cent of their mass due to stellar evolution and dynamical mass loss, the truncation masses are typically higher than present day GC masses.
With the exception of the lowest mass galaxies, where $\Mcstar$ remains approximately constant for $\ZH \lesssim -1$ dex, the truncation mass increases with metallicity with an increase that is steeper in higher mass galaxies. 
The 95th percentile of the truncation mass distribution increases from $1.6 \times 10^{6} \Msun$ for the $<10^{9} \Msun$ galaxies to $2.2 \times 10^{6} \Msun$ for the intermediate mass ($10^{9}$--$10^{10}~\Msun$) galaxies and $3.5 \times 10^{7} \Msun$ for the galaxies more massive than $10^{10} \Msun$.
The 95th percentiles of the truncation mass distribution for star particles with metallicities in the range $-3 <$ [Z/H] $< -2$ are $3.7 \times 10^{6}$, $3.7 \times 10^{6}$ and $6.0 \times 10^{6} \Msun$ respectively.
Therefore, the highest truncation masses are only reached at high metallicities and only in massive galaxies.
The small difference between the upper envelope of the truncation masses at high and low metallicities in low-mass galaxies explains why a much weaker blue tilt is observed in low-mass galaxies than in massive galaxies (see also Figure \ref{fig:galaxy_mass} and Section \ref{sec:colour_mag}).

Along similar lines, we predict a weaker blue tilt at large galactocentric radii for two reasons (see also Figure \ref{fig:tilt_radius} and Section \ref{sec:colour_mag}).
Firstly, the fraction of GCs formed in low-mass galaxies increases with galactocentric distance, resulting in lower truncation masses.
Secondly, the truncation mass for in-situ star formation at large radii is lower \citep{2017MNRAS.469.1282R} since gas surface densities decline with increasing galactocentric radius \citep{2015MNRAS.450.1937C}.
Together, this weakens the blue tilt.
In this context, we note that in all \emosaics simulations, the GC populations of galaxies are made up of a mixture of in-situ formation and ex-situ accretion.
Therefore, the hierarchical build-up of GC populations is explicitly included in the simulations.
In our model, the differences in the blue tilt between galaxies of similar mass is due to differences in their formation and assembly history.

At first sight, our model appears similar to the origin of the blue tilt proposed by \citet{2018arXiv180103515C}, but important physical differences exist between both models.
They both rely on the idea that the conditions for forming the most massive clusters preferentially exist in more massive (and thus more metal-rich) galaxies.
However, the physically motivated E-MOSAICS model for star cluster formation is dependent on the local gas properties and matches star cluster formation in the local universe \citep[Pfeffer et al. in prep.]{2012MNRAS.426.3008K, 2017MNRAS.469.1282R}.
By contrast, the \citet{2018arXiv180103515C} model only triggers star cluster formation during major mergers and assumes that the star cluster properties depend on the integrated gas properties of the host galaxy.
These properties are calculated from galaxy scaling relations and merger trees from a dark matter-only simulation.
In E-MOSAICS, the local gas properties are calculated self-consistently as part of the EAGLE cosmological hydrodynamical galaxy model.
We also note that EAGLE reproduces a wide range of galaxy observables.

In \emosaics, the blue tilt occurs without need the for any additional physics, since the truncation mass of the star cluster initial mass function increases, on average, with metallicity.
This occurs because more massive galaxies tend to have deeper potentials, allowing for higher gas densities, and thus larger truncation masses, and since they have higher metallicities, on average.
Lower mass galaxies and the outskirts of higher mass galaxies should have weaker blue tilts than the centres of massive galaxies, since they have a weaker relation between truncation mass and metallicity.

\subsection{Varying cluster formation physics}
\label{sec:alt_physics}

\begin{figure*}
\begin{center}
\includegraphics[width=504pt]{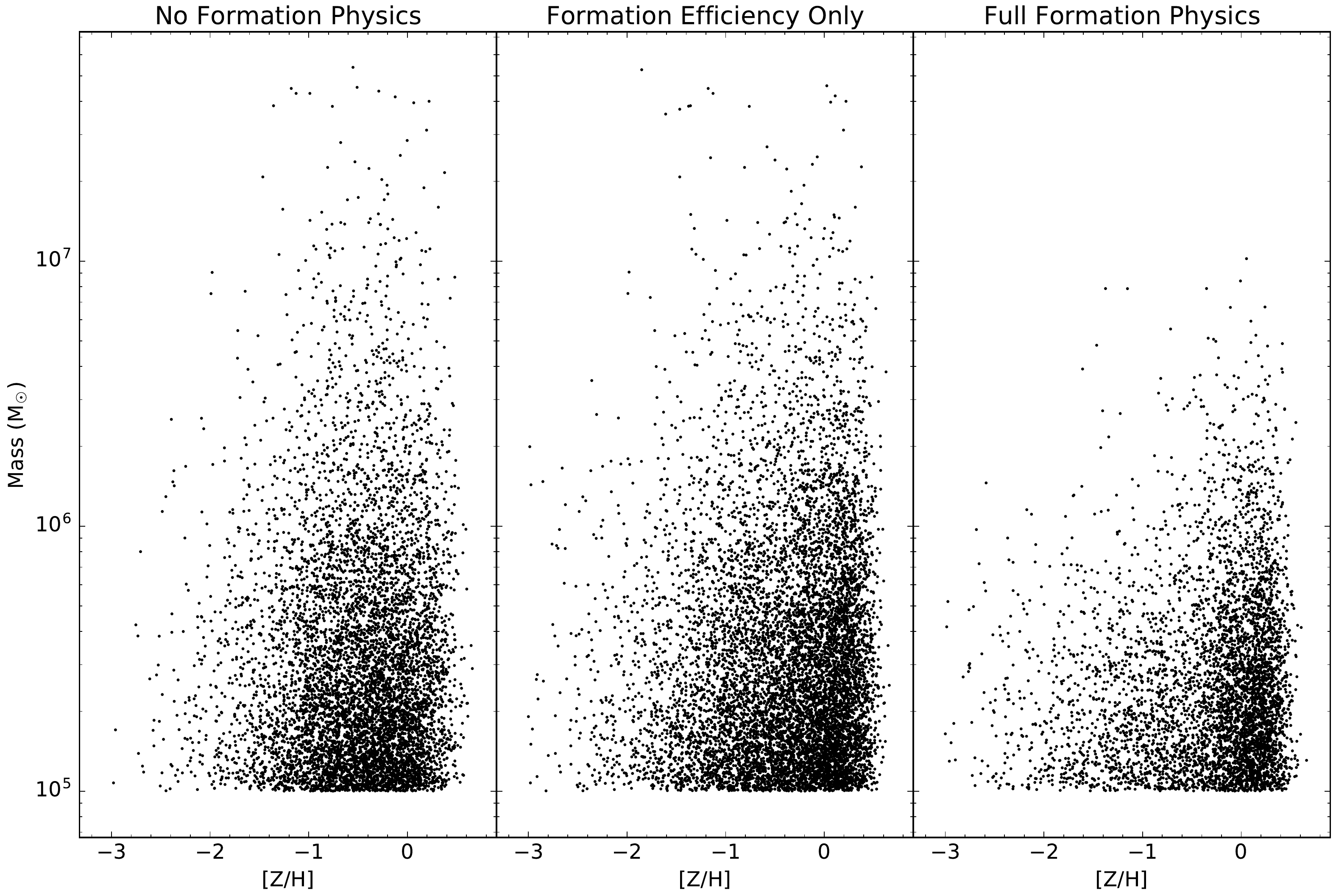}
\caption{Mass-metallicity plots for GCs from simulations with no environmentally dependent star cluster formation physics (left), from simulations with only a cluster formation efficiency model and no maximum cluster mass (centre), and from simulations with the full \emosaics cluster formation physics (right).
Notice the relative lack of massive, metal-poor GCs in the simulations with full GC formation physics compared to simulations with no maximum star cluster mass.
This variation of the truncation mass with metallicity causes the blue tilt in the \emosaics galaxies.
The models without a truncation mass produce significant numbers of GCs more massive than $10^{7}$ M$_{\sun}$.
These are not present in the full formation physics model, which includes a truncation mass.}
\label{fig:formation_mass_metal}
\end{center}
\end{figure*}

A major advantage of the \mosaics approach to modelling star clusters is that different components of the cluster formation and evolution models can be enabled and disabled independently to assess their effect on the final star cluster populations.
To test our model that the blue tilt is caused by a variation in the maximum star cluster formation mass with metallicity, we have also analysed the star cluster populations of the first ten of our zoom in simulations with alternative star cluster formation physics \citep[first described in][]{2018MNRAS.475.4309P}.

In Figure \ref{fig:formation_mass_metal}, we present mass-metallicity plots for GCs from three different simulations: (1) with a constant cluster formation efficiency of 10~per~cent and no maximum mass limit
(`no formation physics' model), (2) with a gas pressure dependent cluster formation efficiency \citep[from][]{2012MNRAS.426.3008K} but no maximum mass limit (`formation efficiency only' model) and (3) with the fiducial \emosaics cluster formation physics (both gas pressure dependent cluster formation efficiency and cluster truncation mass dependent on gas pressure and the coherent motion of the gas via the local epicyclic frequency.)

In both sets of alternative formation physics simulations clusters reach significantly higher masses than in the fiducial model due to the absence of a finite truncation mass.
The alternative models also show evidence for a maximum cluster mass that increases with metallicity, however this is simply caused by the formation of fewer GCs at lower metallicities ($\ZH < -1.5$) than at higher metallicities (i.e.~a size-of-sample effect).
However, as we will find in the following section (see Figure \ref{fig:formation}), both of the alternative formation models result in significantly weaker blue tilts than in the fiducial model, which are too weak to explain blue tilts in observed GC populations.

\section{Measuring the blue tilt as a colour-magnitude relation in E-MOSAICS}
\label{sec:colour_mag}

To quantify the strength of the blue tilt in colour-magnitude space, we perform the same analysis that has been performed in observational studies \citep[e.g.][]{2006ApJ...653..193M, 2009ApJ...699..254H, 2013MNRAS.436.1172U}.
We split the population by absolute magnitude into bins with equal numbers of GCs in each.
As a compromise between the number of magnitude bins and the number GCs within each bin, we calculate the number of bins as $N_{\rm{GC}} / (2.5 \sqrt{N_{\rm{GC}}} + 25)$ where $N_{\rm{GC}}$ is the total number of GCs.
As such we use 37 bins each with 285 GCs for our full sample.
In each bin we fit the colour distribution with a combination of two heteroscedastic Gaussians using the \textsc{scikit-learn} \textsc{GMM} class \citep{scikit-learn} to find the mean colours of the red and blue subpopulations.
We then fit the mean colours of the blue and red Gaussians as linear functions of the mean absolute magnitudes of the GCs in each magnitude bin.
To estimate the uncertainties of our colour-magnitude relations, we perform bootstrapping with 1024 samples.
We project the resulting mean colours and colour-magnitude relations on the colour-magnitude diagram in Figure \ref{fig:all_blue_tilt} for our entire GC sample.
The small pile of up GCs at $(g - z) = 0.84$ is due to the change in slope of the \citet{2012MNRAS.426.1475U} colour-metallicity relation and has no significant effect on the measured blue tilt.
We find a significant ($\sim4\sigma$) blue colour-magnitude relation, with a slope of $d(g - z)/dM_{z} = -0.0279_{-0.0054}^{+0.0069}$, but find only a weak ($\sim2\sigma$) red colour-magnitude relation, with a slope of $d(g - z)/dM_{z} = -0.0049_{-0.0024}^{+0.0020}$.
Our fitted relations are in excellent agreement with those fitted to a combined Virgo and Fornax cluster sample by \citet[see][who find $d(g - z)/dM_{z} =-0.0293 \pm 0.0085$ and $d(g - z)/dM_{z} = -0.0082 \pm 0.0190$, respectively]{2010ApJ...710.1672M}.

Using our adopted colour-metallicity relation, we can convert these colour-magnitude relations into mass-metallicity relations, finding $\gamma = 0.190_{-0.044}^{+0.034}$ and $\gamma = 0.031_{-0.013}^{+0.016}$ for the blue and red subpopulations, respectively, where $Z \propto M^{\gamma}$.
If we fit relations directly to the two Gaussian fits of the metallicity distribution, in mass bins in a similar manner to our colour-magnitude relations, we find consistent mass-metallicity slopes of $\gamma = 0.128_{-0.044}^{+0.068}$ and $\gamma = 0.026_{-0.015}^{+0.024}$, respectively.
We note that care should be taken to compare these mass-metallicity slopes with observational studies as different colour-metallicity relations will yield different mass-metallicity relations from the same colour-magnitude relation.
Our mass-metallicity relation is weaker than that predicted by the models of \citet{2018arXiv180103515C}, although their sample is dominated by galaxies with significantly higher halo masses than ours.
Indeed, our models predict an increase of the strength of the blue tilt towards higher galaxy masses (see below).

We perform the same blue tilt analysis on the 22 individual \emosaics galaxies that have at least 150 GCs satisfying our selection criteria.
We present colour-magnitude diagrams of each of these galaxies in Figure \ref{fig:per_halo_blue_tilt} and give the values of the slopes of the colour-magnitude relations in Table \ref{tab:slopes}.
The 22 galaxies show a diversity of blue tilts with only a few significant `red tilts'.
The range of blue tilt slopes is broadly consistent with those observed \citep[e.g.][]{2010ApJ...710.1672M, 2009ApJ...699..254H} for galaxies of similar masses.

\begin{table*}
\caption{Colour-magnitude relations}
\label{tab:slopes}
\begin{tabular}{lccccccccc}
\hline
Galaxy & Halo mass & Stellar mass & N$_{gc}$ & $d(g - z)_{b}/dM_{z}$ & $\gamma_{b}$ & $p_{b}$ & $d(g - z)_{r}/dM_{z}$ & $\gamma_{r}$ & $p_{r}$ \\
(1) & (2) & (3) & (4) & (5) & (6) & (7) & (8) & (9) & (10) \\
\hline
MW00\_0000   & 12.72 & 10.94 & 895 & $-0.0432_{-0.0203}^{+0.0218}$ & $\hphantom{-}0.277_{-0.139}^{+0.130}$ & 0.033 & $-0.0101_{-0.0110}^{+0.0065}$ & $\hphantom{-}0.065_{-0.042}^{+0.130}$ & 0.077 \\
MW00\_1715   & 12.43 & 10.44 & 243 & $-0.0623_{-0.0337}^{+0.0683}$ & $\hphantom{-}0.399_{-0.437}^{+0.216}$ & 0.190 & $-0.0582_{-0.0113}^{+0.0347}$ & $\hphantom{-}0.373_{-0.222}^{+0.216}$ & 0.023 \\
MW00\_3203 T & 11.95 & 10.32 & 233 & $-0.1142_{-0.0478}^{+0.0657}$ & $\hphantom{-}0.731_{-0.420}^{+0.306}$ & 0.036 & $-0.0163_{-0.0232}^{+0.0292}$ & $\hphantom{-}0.105_{-0.187}^{+0.306}$ & 0.387 \\
MW01\_0000 T & 12.15 & 10.40 & 331 & $\hphantom{-}0.0218_{-0.0774}^{+0.0379}$ & $-0.140_{-0.242}^{+0.496}$ & 0.520 & $\hphantom{-}0.0056_{-0.0163}^{+0.0039}$ & $-0.036_{-0.025}^{+0.496}$ & 0.491 \\
MW02\_0000 T & 12.32 & 10.61 & 748 & $-0.0505_{-0.0272}^{+0.0317}$ & $\hphantom{-}0.323_{-0.203}^{+0.174}$ & 0.052 & $\hphantom{-}0.0021_{-0.0074}^{+0.0053}$ & $-0.013_{-0.034}^{+0.174}$ & 0.551 \\
MW03\_0329 T & 12.17 & 10.45 & 474 & $-0.0382_{-0.0614}^{+0.0366}$ & $\hphantom{-}0.245_{-0.234}^{+0.393}$ & 0.152 & $-0.0187_{-0.0156}^{+0.0059}$ & $\hphantom{-}0.120_{-0.038}^{+0.393}$ & 0.014 \\
MW04\_0000 T & 12.03 & 10.15 & 206 & $\hphantom{-}0.0304_{-0.0577}^{+0.1201}$ & $-0.194_{-0.768}^{+0.370}$ & 0.706 & $\hphantom{-}0.0031_{-0.0171}^{+0.0325}$ & $-0.020_{-0.208}^{+0.370}$ & 0.667 \\
MW05\_0457   & 12.18 & 10.41 & 291 & $\hphantom{-}0.0055_{-0.0430}^{+0.0749}$ & $-0.035_{-0.480}^{+0.275}$ & 0.607 & $\hphantom{-}0.0084_{-0.0116}^{+0.0218}$ & $-0.054_{-0.139}^{+0.275}$ & 0.790 \\
MW05\_0766 T & 12.09 & 10.15 & 852 & $-0.0515_{-0.0246}^{+0.0542}$ & $\hphantom{-}0.330_{-0.347}^{+0.157}$ & 0.167 & $-0.0043_{-0.0035}^{+0.0059}$ & $\hphantom{-}0.028_{-0.038}^{+0.157}$ & 0.235 \\
MW06\_0000 T & 11.98 & 10.32 & 302 & $\hphantom{-}0.0713_{-0.0399}^{+0.0355}$ & $-0.456_{-0.227}^{+0.255}$ & 0.954 & $\hphantom{-}0.0170_{-0.0102}^{+0.0056}$ & $-0.109_{-0.036}^{+0.255}$ & 0.964 \\
MW09\_0000 T & 11.90 & 10.19 & 166 & $\hphantom{-}0.0356_{-0.0637}^{+0.0976}$ & $-0.228_{-0.624}^{+0.408}$ & 0.747 & $\hphantom{-}0.0188_{-0.0174}^{+0.0187}$ & $-0.120_{-0.120}^{+0.408}$ & 0.856 \\ 
MW10\_0000 T & 12.46 & 10.54 & 479 & $-0.0155_{-0.0328}^{+0.0443}$ & $\hphantom{-}0.099_{-0.283}^{+0.210}$ & 0.369 & $-0.0143_{-0.0071}^{+0.0110}$ & $\hphantom{-}0.091_{-0.070}^{+0.210}$ & 0.096 \\
MW12\_0000 T & 12.39 & 10.50 & 743 & $-0.0543_{-0.0166}^{+0.0254}$ & $\hphantom{-}0.348_{-0.163}^{+0.106}$ & 0.041 & $-0.0059_{-0.0046}^{+0.0063}$ & $\hphantom{-}0.038_{-0.040}^{+0.106}$ & 0.181 \\
MW12\_0001   & 11.80 & 10.21 & 403 & $\hphantom{-}0.0579_{-0.0466}^{+0.0337}$ & $-0.370_{-0.216}^{+0.298}$ & 0.899 & $-0.0002_{-0.0065}^{+0.0059}$ & $\hphantom{-}0.001_{-0.038}^{+0.298}$ & 0.485 \\
MW12\_1009   & 12.16 & 10.41 & 226 & $-0.0387_{-0.0413}^{+0.0630}$ & $\hphantom{-}0.247_{-0.403}^{+0.264}$ & 0.296 & $\hphantom{-}0.0173_{-0.0173}^{+0.0156}$ & $-0.111_{-0.100}^{+0.264}$ & 0.840 \\
MW13\_0000 T & 12.40 & 10.39 & 170 & $\hphantom{-}0.0043_{-0.0527}^{+0.0530}$ & $-0.028_{-0.339}^{+0.338}$ & 0.554 & $-0.0025_{-0.0257}^{+0.0353}$ & $\hphantom{-}0.016_{-0.226}^{+0.338}$ & 0.531 \\
MW14\_0000 T & 12.44 & 10.64 & 238 & $-0.0776_{-0.0475}^{+0.0361}$ & $\hphantom{-}0.497_{-0.231}^{+0.304}$ & 0.031 & $-0.0017_{-0.0146}^{+0.0129}$ & $\hphantom{-}0.011_{-0.083}^{+0.304}$ & 0.423 \\
MW16\_0000 T & 12.36 & 10.62 & 444 & $-0.0575_{-0.0282}^{+0.0400}$ & $\hphantom{-}0.368_{-0.256}^{+0.181}$ & 0.064 & $-0.0310_{-0.0127}^{+0.0088}$ & $\hphantom{-}0.199_{-0.057}^{+0.181}$ & 0.001 \\
MW21\_1233 T & 12.18 & 10.11 & 151 & $\hphantom{-}0.0189_{-0.0412}^{+0.0316}$ & $-0.121_{-0.202}^{+0.264}$ & 0.646 & $-0.0465_{-0.0192}^{+0.0321}$ & $\hphantom{-}0.297_{-0.206}^{+0.264}$ & 0.075 \\
MW22\_0515 T & 12.17 & 10.48 & 264 & $-0.0455_{-0.0271}^{+0.0379}$ & $\hphantom{-}0.291_{-0.243}^{+0.173}$ & 0.125 & $\hphantom{-}0.0102_{-0.0110}^{+0.0118}$ & $-0.065_{-0.075}^{+0.173}$ & 0.824 \\
MW23\_0000 T & 12.27 & 10.58 & 388 & $-0.0941_{-0.0093}^{+0.1356}$ & $\hphantom{-}0.602_{-0.868}^{+0.060}$ & 0.283 & $\hphantom{-}0.0040_{-0.0076}^{+0.0163}$ & $-0.026_{-0.104}^{+0.060}$ & 0.755 \\
MW23\_0001   & 11.84 & 10.09 & 219 & $-0.0260_{-0.0562}^{+0.0290}$ & $\hphantom{-}0.167_{-0.186}^{+0.359}$ & 0.185 & $\hphantom{-}0.0241_{-0.0258}^{+0.0091}$ & $-0.154_{-0.058}^{+0.359}$ & 0.806 \\ \hline
Main Sample &  &  & 10553 & $-0.0297_{-0.0054}^{+0.0069}$ & $\hphantom{-}0.190_{-0.044}^{+0.034}$ & 0.000 & $-0.0049_{-0.0024}^{+0.0020}$ & $\hphantom{-}0.031_{-0.013}^{+0.016}$ & 0.010 \\ \hline
\multicolumn{3}{l}{Host galaxy $M_{*} < 1 \times 10^{10}$}   & 764 & $\hphantom{-}0.0051_{-0.0202}^{+0.018}$ & $-0.033_{-0.115}^{+0.129}$ & 0.597 & $-0.0053_{-0.0253}^{+0.0215}$ & $\hphantom{-}0.034_{-0.138}^{+0.162}$ & 0.395 \\
\multicolumn{3}{l}{Host galaxy $1 \times 10^{10} < M_{*} < 3 \times 10^{10}$}   & 5742 & $-0.0203_{-0.0093}^{+0.0097}$ & $\hphantom{-}0.130_{-0.062}^{+0.060}$ & 0.022 & $-0.0007_{-0.0029}^{+0.0025}$ & $\hphantom{-}0.004_{-0.016}^{+0.019}$ & 0.373 \\
\multicolumn{3}{l}{Host galaxy $M_{*} > 3 \times 10^{10}$}   & 4047 & $-0.0342_{-0.0121}^{+0.0134}$ & $\hphantom{-}0.219_{-0.086}^{+0.077}$ & 0.019 & $-0.0054_{-0.0038}^{+0.0038}$ & $\hphantom{-}0.034_{-0.025}^{+0.024}$ & 0.074 \\ \hline
$r_{gc} < 6.7$ kpc &  &  & 2314 & $-0.0358_{-0.0111}^{+0.0213}$ & $\hphantom{-}0.229_{-0.137}^{+0.071}$ & 0.042 & $-0.0022_{-0.0042}^{+0.0048}$ & $\hphantom{-}0.014_{-0.031}^{+0.027}$ & 0.340 \\
$r_{gc} > 6.7$ kpc &  &  & 1733 & $-0.0206_{-0.0111}^{+0.0146}$ & $\hphantom{-}0.132_{-0.094}^{+0.071}$ & 0.077 & $-0.0126_{-0.0065}^{+0.0110}$ & $\hphantom{-}0.080_{-0.070}^{+0.042}$ & 0.117 \\ 
$R_{XY} < 5$ kpc &  &  & 2300 & $-0.0351_{-0.0100}^{+0.0237}$ & $\hphantom{-}0.225_{-0.152}^{+0.064}$ & 0.060 & $-0.0028_{-0.0038}^{+0.0053}$ & $\hphantom{-}0.018_{-0.034}^{+0.024}$ & 0.313 \\
$R_{XY} > 5$ kpc &  &  & 1747 & $-0.0271_{-0.0080}^{+0.0198}$ & $\hphantom{-}0.173_{-0.126}^{+0.051}$ & 0.066 & $-0.0126_{-0.0063}^{+0.0102}$ & $\hphantom{-}0.081_{-0.065}^{+0.040}$ & 0.096 \\
$R_{XZ} < 5$ kpc &  &  & 2312 & $-0.0328_{-0.0142}^{+0.0191}$ & $\hphantom{-}0.210_{-0.122}^{+0.091}$ & 0.039 & $-0.0013_{-0.0043}^{+0.0042}$ & $\hphantom{-}0.008_{-0.027}^{+0.028}$ & 0.389 \\
$R_{XZ} > 5$ kpc &  &  & 1735 & $-0.0224_{-0.0100}^{+0.0162}$ & $\hphantom{-}0.143_{-0.104}^{+0.064}$ & 0.078 & $-0.0134_{-0.0065}^{+0.0107}$ & $\hphantom{-}0.085_{-0.069}^{+0.042}$ & 0.114 \\
$R_{YZ} < 5$ kpc &  &  & 2278 & $-0.0393_{-0.0090}^{+0.0240}$ & $\hphantom{-}0.251_{-0.154}^{+0.058}$ & 0.036 & $-0.0025_{-0.0037}^{+0.0051}$ & $\hphantom{-}0.016_{-0.033}^{+0.024}$ & 0.336 \\
$R_{YZ} > 5$ kpc &  &  & 1769 & $-0.0181_{-0.0141}^{+0.0117}$ & $\hphantom{-}0.116_{-0.075}^{+0.090}$ & 0.067 & $-0.0237_{-0.0069}^{+0.0101}$ & $\hphantom{-}0.152_{-0.065}^{+0.044}$ & 0.007 \\ \hline
\multicolumn{3}{l}{No Formation Physics} & 7348 & $-0.0120_{-0.0035}^{+0.0040}$ & $\hphantom{-}0.077_{-0.026}^{+0.023}$ & 0.001 & $-0.0045_{-0.0026}^{+0.0032}$ & $\hphantom{-}0.029_{-0.020}^{+0.017}$ & 0.064 \\
\multicolumn{3}{l}{Formation Efficiency Only} & 9394 & $-0.0124_{-0.0034}^{+0.0045}$ & $\hphantom{-}0.079_{-0.029}^{+0.022}$ & 0.003 & $-0.0096_{-0.0026}^{+0.0025}$ & $\hphantom{-}0.062_{-0.016}^{+0.017}$ & 0.001 \\
\multicolumn{3}{l}{Full Formation Physics} & 5138 & $-0.0316_{-0.0141}^{+0.0091}$ & $\hphantom{-}0.202_{-0.058}^{+0.090}$ & 0.007 & $-0.0020_{-0.0040}^{+0.0023}$ & $\hphantom{-}0.013_{-0.014}^{+0.026}$ & 0.176 \\
 \hline
\end{tabular}
%\hphantom{-}
\medskip
\emph{Notes}
Column (1): Galaxy identifier or sample name. The Galaxy identifiers are of the form MWNN\_MMMM where NN is the zoom number and MMMM is the subhalo identifier. T indicates the galaxies that the zooms targeted.
Column (2): Galaxy halo mass in solar masses.
Column (3): Galaxy stellar mass in solar masses.
Column (4): Number of selected GCs in galaxy or sample.
Column (5): Slope of the colour-magnitude relation for the blue subpopulation.
Column (6): Power law index of the mass-metallicity relation ($Z \propto M^{\gamma}$) calculated from the slope of the blue colour-magnitude relation.
Column (7): Bootstrap probability that the blue colour-magnitude slope is greater than zero.
Column (8): Slope of the colour-magnitude relation for the red subpopulation.
Column (9): Power law index of the mass-metallicity relation calculated from the slope of the red colour-magnitude relation.
Column (10): Bootstrap probability that the red colour-magnitude slope is greater than zero.
\end{table*}

\begin{figure*}
\begin{center}
\includegraphics[width=504pt]{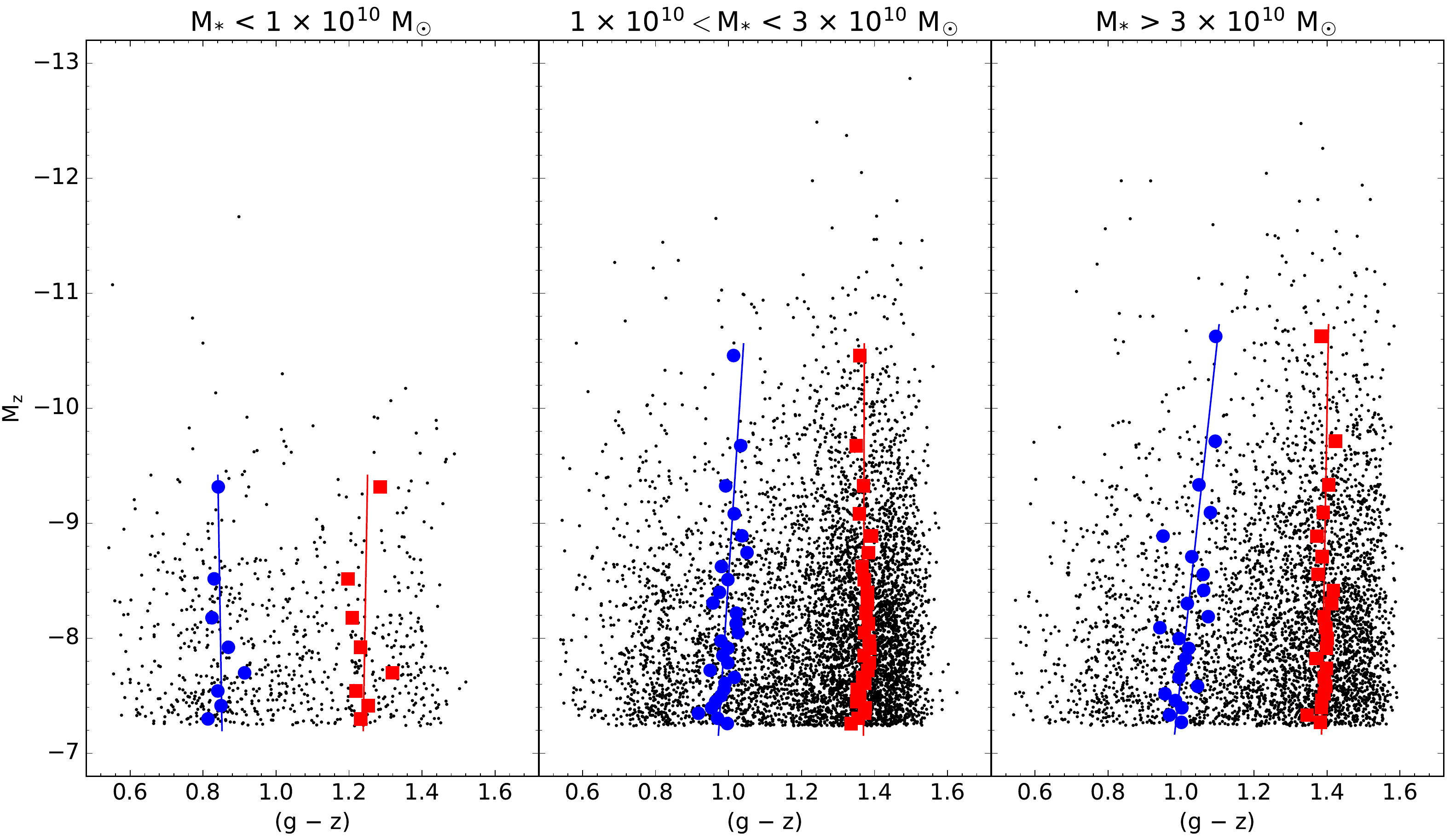}
\caption{Blue tilt of GCs in host galaxies with stellar masses $M_*<1\times10^{10}$ M$_{\sun}$ (left), $1 \times 10^{10}<M_*/\Msun<3 \times 10^{10}$ (centre), and $M_*>3 \times 10^{10}$ M$_{\sun}$ (right).
The lowest-mass galaxies in our sample (left-hand panel) show no evidence for a blue tilt, whereas the highest mass galaxies (right-hand panel) show a stronger blue tilt than the intermediate mass galaxies (central panel).}
\label{fig:galaxy_mass}
\end{center}
\end{figure*}

In Figure \ref{fig:galaxy_mass}, we investigate the effects of galaxy mass on the blue tilt by splitting our sample into three host galaxy stellar masses bins.
Our galaxy bins are $M_*<1\times10^{10}$ M$_{\sun}$,  $1 \times 10^{10}<M_*/\Msun<3 \times 10^{10}$, and $M_*>3 \times 10^{10}$.
Using the same analysis as for the combined sample and for the individual galaxies, we find slopes of $d(g - z) / dM_{z} = 0.0051_{-0.0054}^{+0.0069}$, $-0.0203_{-0.0093}^{+0.0097}$, and $-0.0342_{-0.0130}^{+0.0122}$, for the increasing galaxy mass bins.
This is compatible with the observed weakening of the blue tilt at lower galaxy luminosities \citep{2006ApJ...653..193M, 2010ApJ...710.1672M} and with how the relationship between truncation mass and metallicity varies with galaxy mass (Figure \ref{fig:mcstar}).

\begin{figure*}
\begin{center}
\includegraphics[width=504pt]{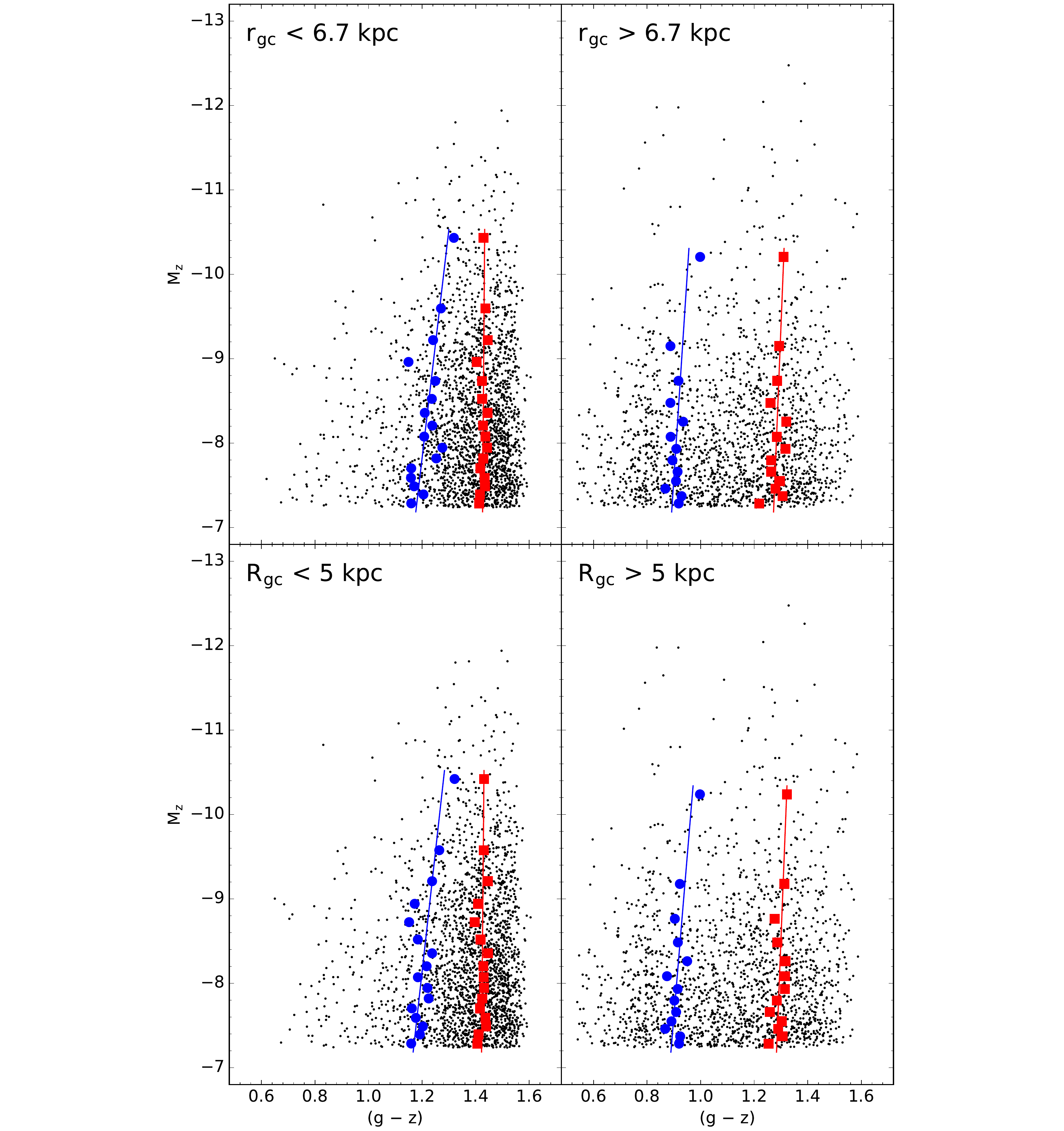}
\caption{Blue tilt of GCs within a three-dimensional galactocentric radius of 6.7 kpc (top left) and beyond 6.7 kpc (top right), as well as of GCs within a projected radius of 5 kpc (bottom left) and beyond 5 kpc (bottom right).
Both the projected and physically inner GCs show slighter stronger blue tilts compared to the outer GCs in line with observations.}
\label{fig:tilt_radius}
\end{center}
\end{figure*}

We also investigate how the blue tilt changes with distance from the centre of the galaxy by considering only those galaxies with stellar mass $M_*>3 \times 10^{10}$ M$_{\sun}$.
Firstly, in the bottom half of Figure \ref{fig:tilt_radius} we split the sample of GCs by their 3-dimensional galactocentric distance.
The GCs within 6.7 kpc show a slightly stronger blue tilt compared to the GCs beyond 6.7 kpc ($d(g - z) / dM_{z} = -0.0358_{-0.0111}^{+0.0213}$ versus $d(g - z) / dM_{z} = -0.0206_{-0.0111}^{+0.0146}$), with the inner GCs showing a similar slope to that for all GCs in galaxies more massive than $3 \times 10^{10}$ M$_{\sun}$ ($-0.0342_{-0.0130}^{+0.0122}$).
Secondly, in the top of Figure \ref{fig:tilt_radius} we split the sample of GCs by their galactocentric distance projected in the $XY$-plane of the simulations.
Observational studies beyond the Local Group can only measure the projected distance between a GC and the centre of its host galaxy.
We find both GCs within and beyond 5 kpc in projection show a blue tilt and that there is little evidence that the central GCs have a stronger blue tilt in this projection ($d(g - z) / dM_{z} = -0.0351_{-0.0100}^{+0.0237}$ versus $d(g - z) / dM_{z} = -0.0271_{-0.0080}^{+0.0198}$).
We see similar inner and outer blue tilts for the $XZ$ and $YZ$ projections (see Table \ref{tab:slopes}). 
We note that our galaxies have random orientations with respect to the coordinate system of the simulations.
To test what effect projection angle has on our measured blue tilt, we rotated our models by a random angle 3072 times and calculated the projected blue tilt for each rotation.
In projection, the mean inner blue tilt has a slope of $d(g - z) / dM_{z} = -0.0347_{-0.0091}^{+0.110}$ and the mean outer blue tilt has a slope of $d(g - z) / dM_{z} = -0.0218_{-0.0039}^{+0.0042}$.
These stronger blue tilts in the centres of galaxies are in line with most \citep[e.g.][]{2006ApJ...653..193M, 2010ApJ...710.1672M, 2013MNRAS.436.1172U} but not all observations \citep{2009ApJ...699..254H} although these studies are of more massive galaxies than are considered here (their stellar masses $\gtrsim 10^{11} \Msun$).
A weaker blue tilt at larger radii is expected due to less variation in truncation mass with metallicity at larger galactocentric distance as GCs at larger radii are more likely to have either been accreted from lower mass galaxies or to have formed in-situ from lower density gas than in the galaxy centre.

We expect that the blue tilt should vary significantly at fixed galaxy mass due to differences in galaxy formation and assembly histories.
Statistically significant, physical correlations likely exist between the slope of the blue tilt and these formation and assembly histories of the host \citep[cf.][]{K18}.
This is an interesting avenue for future work.
Unfortunately, such an effort will be hampered by the stochasticity that affects blue tilt measurements for the galaxy mass range currently covered by \emosaics, due to the relatively small number of metal-poor GCs per galaxy (see Table~\ref{tab:slopes}, Figure~\ref{fig:per_halo_blue_tilt}).
In the future, we aim to extend our analysis to higher halo masses and to a larger sample of galaxies to alleviate this problem.

\begin{figure*}
\begin{center}
\includegraphics[width=504pt]{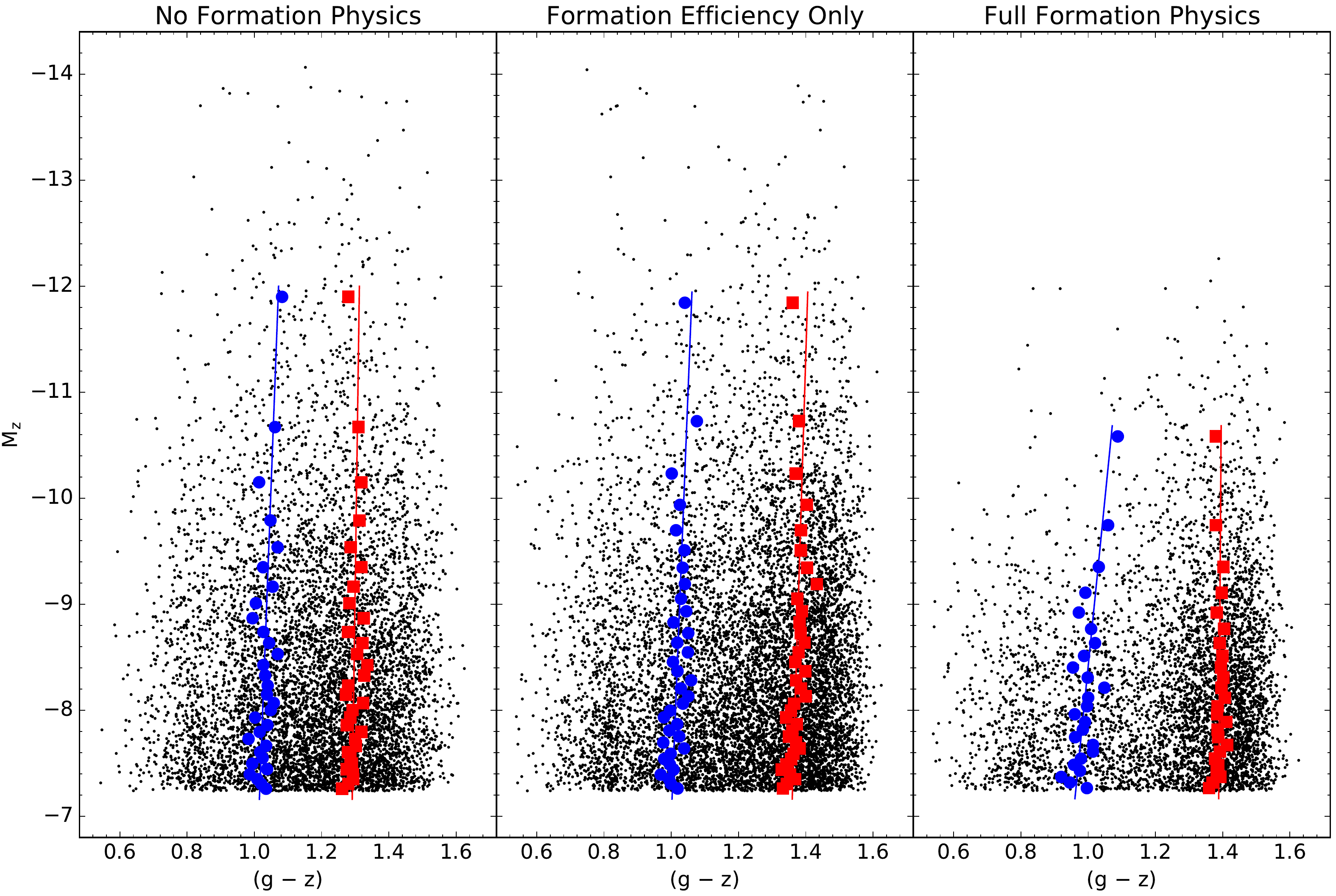}
\caption{Blue tilt of GCs of from simulations with no star cluster formation physics (left), from simulations with only varying cluster formation efficiency and no maximum cluster mass (centre) and from simulations with the full \emosaics cluster physics (right) in colour-magnitude space.
The simulations without a star cluster truncation mass show much weaker blue tilts compared to the simulations with a truncation mass that depends on the local environment (i.e.~gas pressure and centrifugal forces).}
\label{fig:formation}
\end{center}
\end{figure*}

In Figure \ref{fig:formation} we show the version of Figure \ref{fig:formation_mass_metal} in colour-magnitude space for three different simulations (1) with a constant cluster formation efficiency and no maximum mass limit, (2) with a gas pressure dependent cluster formation efficiency but no maximum mass limit and (3) for simulations with the fiducial \emosaics formation physics.
The simulations without a star cluster truncation mass show much weaker blue tilts ($d(g - z) / dM_{z} = -0.0120_{-0.0035}^{+0.0040}$ and $d(g - z) / dM_{z} = -0.0124_{-0.0034}^{+0.0045}$) compared to the simulations with a truncation mass dependent on the gas properties ($d(g - z) / dM_{z} = -0.0316_{-0.0141}^{+0.0091}$).

Although we have studied the blue tilt as a variation in the colour distribution with luminosity for consistency with previous work, we encourage the study of the blue tilt as a variation in the luminosity function with colour, because we find that the blue tilt is caused by a variation in the truncation mass with metallicity.
This is a fundamentally different approach to the blue tilt phenomenon.
Studying the behaviour of the blue tilt via the luminosity function rather than colour distribution should be less sensitive to the relationship between colour and metallicity. 

\section{Conclusion}
\label{sec:conclusion}

The `blue tilt' corresponds to a relative lack of the most massive ($> 10^{6}$ M$_{\sun}$), metal-poor GCs ([Z/H] $\sim -2$) and is commonly observed as a colour-magnitude relation in the blue subpopulation of rich GC systems.
While several models have been proposed to explain the origin of the blue tilt, each of these models faces significant challenges (see Section~\ref{sec:old_theories}).
In particular, the popular self-enrichment model of the blue tilt \citep[e.g.][]{2009ApJ...695.1082B}, in which more massive GCs are posited to incorporate metals synthesised by their massive stars into later generations of star formation, suffers from the critical problem that no extended star formation is observed in young massive star clusters with properties similar to GCs.

In Section \ref{sec:emosaics}, we have shown that in the \emosaics suite of cosmological simulations of GC system formation and evolution \citep{2018MNRAS.475.4309P, K18} the blue tilt naturally arises as a lack of metal-poor GCs at high masses (Figure~\ref{fig:all_blue_tilt}), due to a physical upper cluster mass scale that increases with increasing metallicity in massive galaxies.
Following \citet{2017MNRAS.469.1282R}, the truncation mass increases with galaxy mass and metallicity due to the higher gas pressures (surface densities) for star formation attained in the deeper potentials of more massive galaxies, which in turn results in a blue tilt of their GC populations (Figure~\ref{fig:mcstar}).
In this model, different galaxy formation and assembly histories drive variations in the relationship between truncation mass and metallicity, thus leading to different blue tilts.
To compare the predictions of \emosaics with observations, we use an empirical colour-metallicity relation and a constant mass-to-light ratio. 
Performing the same analysis used in observational studies of the blue tilt, we find colour-magnitude relations of similar strength to those found for the GC populations of observed galaxies.

Like observed galaxies, the \emosaics galaxies show a diverse range of blue tilts (Figure \ref{fig:per_halo_blue_tilt}).
We find no evidence for a significant blue tilt in the \emosaics galaxies with stellar masses $M_*<1 \times 10^{10}$ M$_{\sun}$. Galaxies with masses $M_*>3 \times 10^{10}$ M$_{\sun}$ have stronger blue tilts, on average, than galaxies with intermediate masses ($1 \times 10^{10}<M_*/\Msun<3 \times 10^{10}$, see Figure \ref{fig:galaxy_mass}), in line with observed weakening of the blue tilt towards low galaxy luminosities.
Additionally, we find a slightly stronger blue tilt at smaller galactocentric distance, again consistent with observations (Figure \ref{fig:tilt_radius}).
By switching off several elements of the star cluster formation physics in \emosaics, we have shown that the blue tilt arises due to variations in the maximum star cluster mass with formation conditions (Figure \ref{fig:formation}).
This is broadly consistent with the semi-analytic model of \citet{2018arXiv180103515C}, as both models produce a blue tilt as a consequence of the conditions to form the most massive GCs preferentially existing in more massive, more metal rich galaxies.
As blue tilt is caused by an increase in maximum GC mass with metallicity, it is better to think of the blue tilt as a change in the mass function with metallicity (or observationally a change in the luminosity function with colour) than a change in the metallicity (colour) distribution with mass (luminosity).

Finally, we emphasise that \emosaics reproduces the blue tilt without including any non-standard physics, such as self-enrichment within massive clusters.
Therefore, we conclude that no such mechanisms are required to explain the blue tilt.
Instead, we propose that the existence of the blue tilt is a natural consequence of standard cluster formation physics operating in the evolving interstellar conditions fostered by young, hierarchically-assembling galaxies.

\section*{Acknowledgements}
We wish to thank the referee for their useful comments and suggestions which greatly helped to improve this paper.
CU, JP and NB gratefully acknowledge financial support from the European Research Council (ERC-CoG-646928, Multi-Pop).
RAC and NB gratefully acknowledge financial support from the Royal Society (University Research Fellowships).
JMDK gratefully acknowledges funding from the German Research Foundation (DFG) in the form of an Emmy Noether Research Group (grant number KR4801/1-1, PI Kruijssen).
JMDK and MRC gratefully acknowledge funding from the European Research Council (ERC) under the European Union's Horizon 2020 research and innovation programme via the ERC Starting Grant MUSTANG (grant agreement number 714907, PI Kruijssen).
MRC is supported by a Fellowship from the International Max Planck Research School for Astronomy and Cosmic Physics at the University of Heidelberg (IMPRS-HD).

The study made use of high performance computing facilities at Liverpool John Moores University, partly funded by the Royal Society and LJMU's Faculty of Engineering and Technology, and the DiRAC Data Centric system at Durham University, 
operated by the Institute for Computational Cosmology on behalf of the 
STFC DiRAC HPC Facility (www.dirac.ac.uk). This equipment was funded by 
BIS National E-infrastructure capital grant ST/K00042X/1, STFC capital 
grants ST/H008519/1 and ST/K00087X/1, STFC DiRAC Operations grant 
ST/K003267/1 and Durham University. DiRAC is part of the National 
E-Infrastructure. 

This work made use of the Python packages \textsc{NumPy} \citep{numpy}, \textsc{Scipy} \citep{scipy}, \textsc{scikit-learn} \citep{scikit-learn}, and \textsc{matplotlib} \citep{Matplotlib} as well as \textsc{Astropy}, a community-developed core Python package for astronomy \citep{2013A&A...558A..33A}.

\bibliographystyle{mnras}
\bibliography{bib}{}

\appendix
\section{Effects of different metallicity-to-colour conversions on the blue tilt}
\label{sec:various_cmrs}

\begin{figure*}
\begin{center}
\includegraphics[width=504pt]{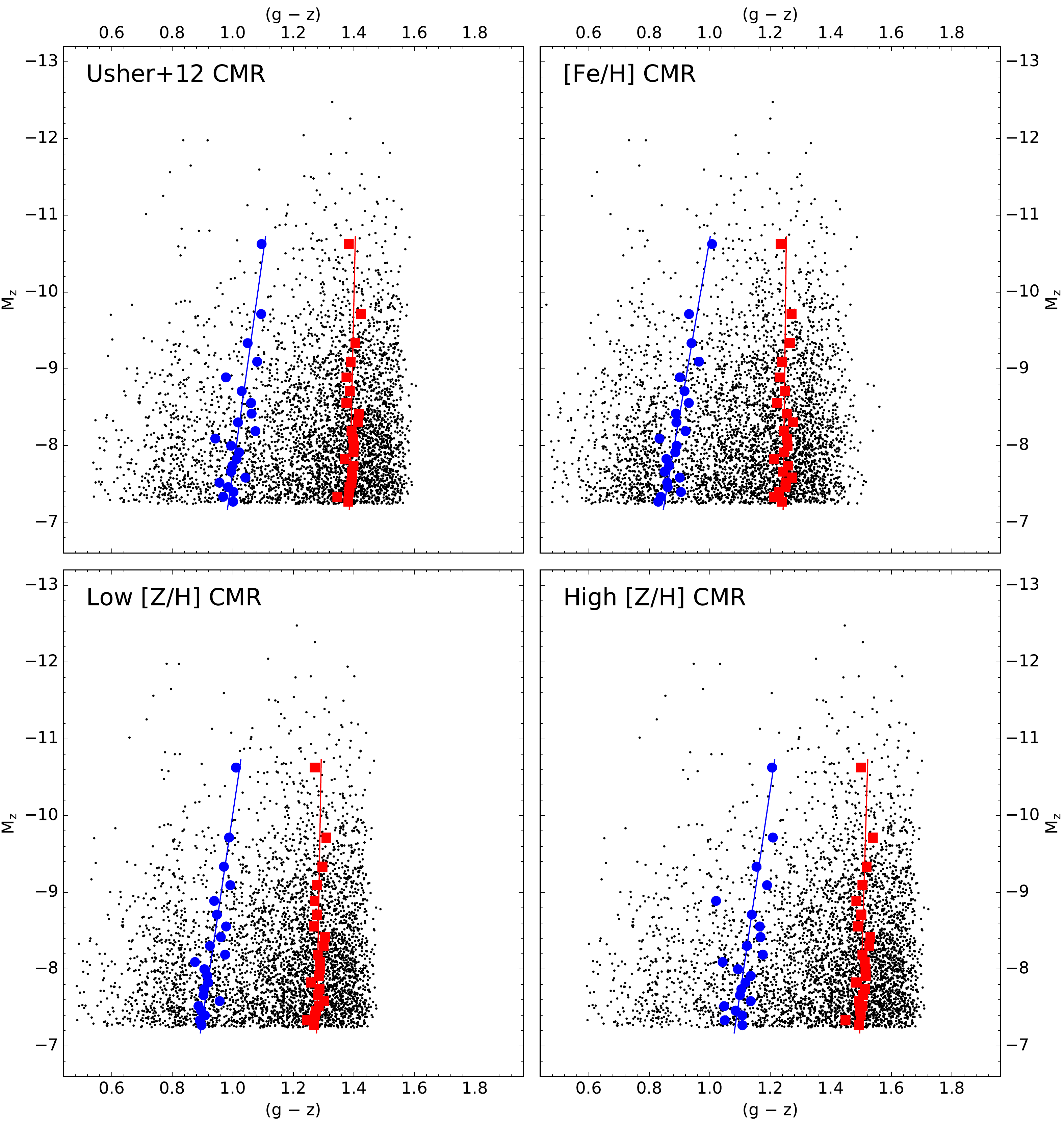}
\caption{Effects of different metallicities on the blue tilt.
\emph{Top left}: Colours calculated from the default \emosaics metallicities using the \citet{2012MNRAS.426.1475U} colour-metallicity relation adopted in this work.
\emph{Top right}: Colours calculated from [Fe/H] rather than [Z/H].
\emph{Bottom left}: Colours calculated from [Z/H] shifted by $-0.3$.
\emph{Bottom right}: Colours calculated from [Z/H] shifted by $+0.3$.
Calculating colours as a function of [Fe/H] rather than [Z/H] or shifting the metallicity by $\pm 0.3$ dex have no significant effect on the measured blue tilt.
The range of colours and absolute magnitudes are the same in each panel.}
\label{fig:cmr_tests_metal}
\end{center}
\end{figure*}

\begin{figure*}
\begin{center}
\includegraphics[width=504pt]{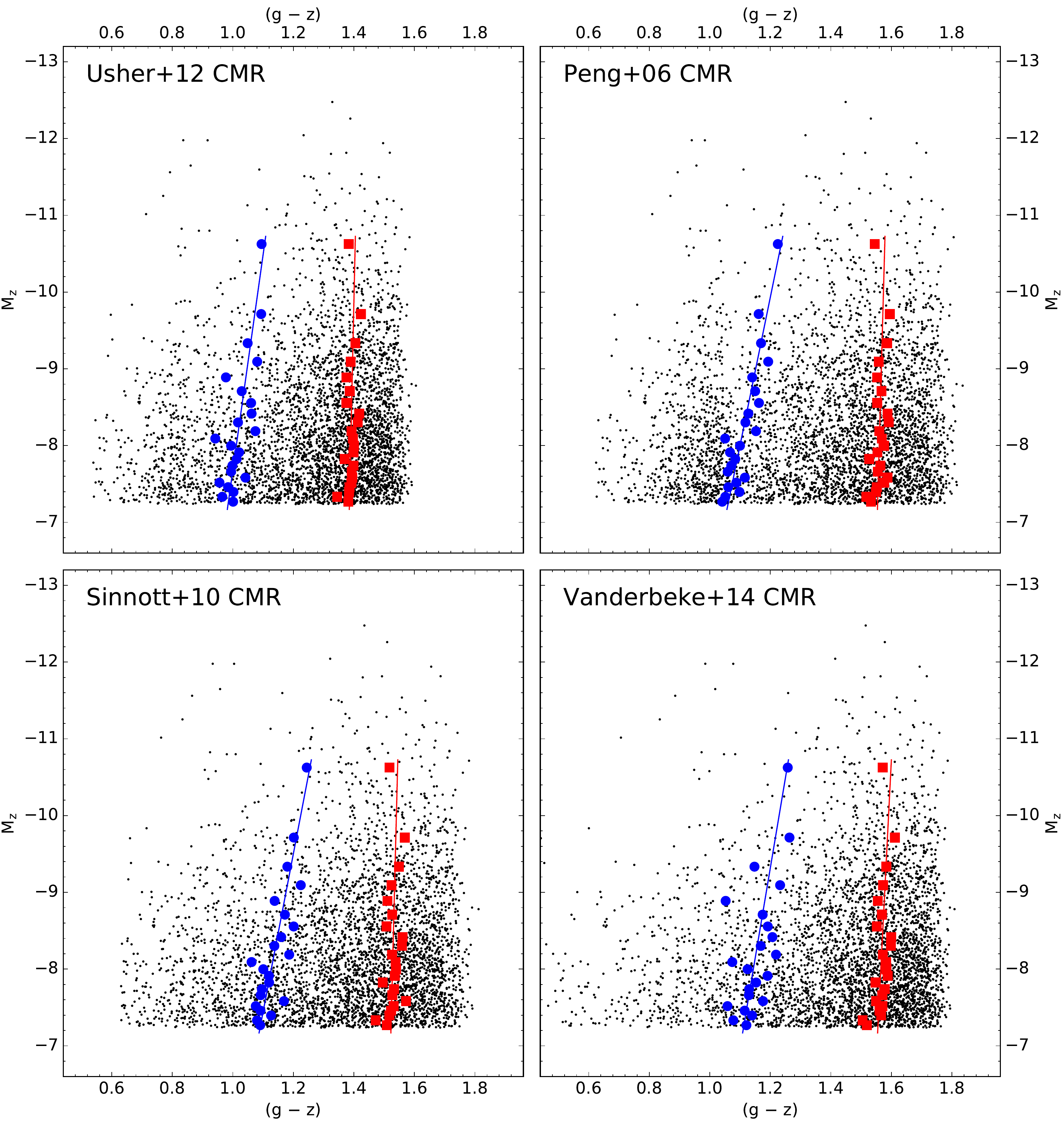}
\caption{Effects of different empirical colour-metallicity relations on the blue tilt.
\emph{Top left}: Colours calculated using the \citet{2012MNRAS.426.1475U} colour-metallicity relation adopted in this work.
\emph{Top right}: Colours calculated using the \citet{2006ApJ...639...95P} colour-metallicity relation.
\emph{Bottom left}: Colours calculated using the \citet{2010AJ....140.2101S} colour-metallicity relation.
\emph{Bottom right}: Colours calculated using the \citet{2014MNRAS.437.1734V} colour-metallicity relation.
While different colour-metallicity relations give different colour-magnitude relation slopes, the blue tilt is qualitatively the same.
The range of colours and absolute magnitudes are the same in each panel and as in Figure \ref{fig:cmr_tests_metal}.}
\label{fig:cmr_tests_cmr}
\end{center}
\end{figure*}

To verify whether our choice of a constant mass-to-light ratio and the \citet{2012MNRAS.426.1475U} empirical colour-metallicity relation have an effect on our results, we repeat our blue tilt analysis on GCs from galaxies more massive than $3 \times 10^{10} \Msun$ with different metallicity-to-colour and mass-to-luminosity conversions.
Firstly, in Figure~\ref{fig:cmr_tests_metal} we show using the iron abundance ([Fe/H]) rather than the metallicity ([Z/H]) as well as testing the effects of shifting the \emosaics metallicities up and down by 0.3 dex to simulate the effects of a systematic offset between observed and simulated metallicities.
None of these changes results in a significantly different blue tilt.
Converting the colour-magnitude relations in to mass-metallicity relations using the slopes of the colour-metallicity relation at the average colours of the subpopulations, we find the relations are all consistent with the relations fitted directly to the masses and metallicities ($0.253_{-0.094}^{+0.082}$ and $0.071_{-0.039}^{+0.026}$ for the metal poor and rich GCs respectively).
Secondly, in Figure~\ref{fig:cmr_tests_cmr} we use three other empirical GC colour-metallicity relations \citep{2006ApJ...639...95P, 2010AJ....140.2101S, 2014MNRAS.437.1734V} to calculate colours.

\begin{figure*}
\begin{center}
\includegraphics[width=504pt]{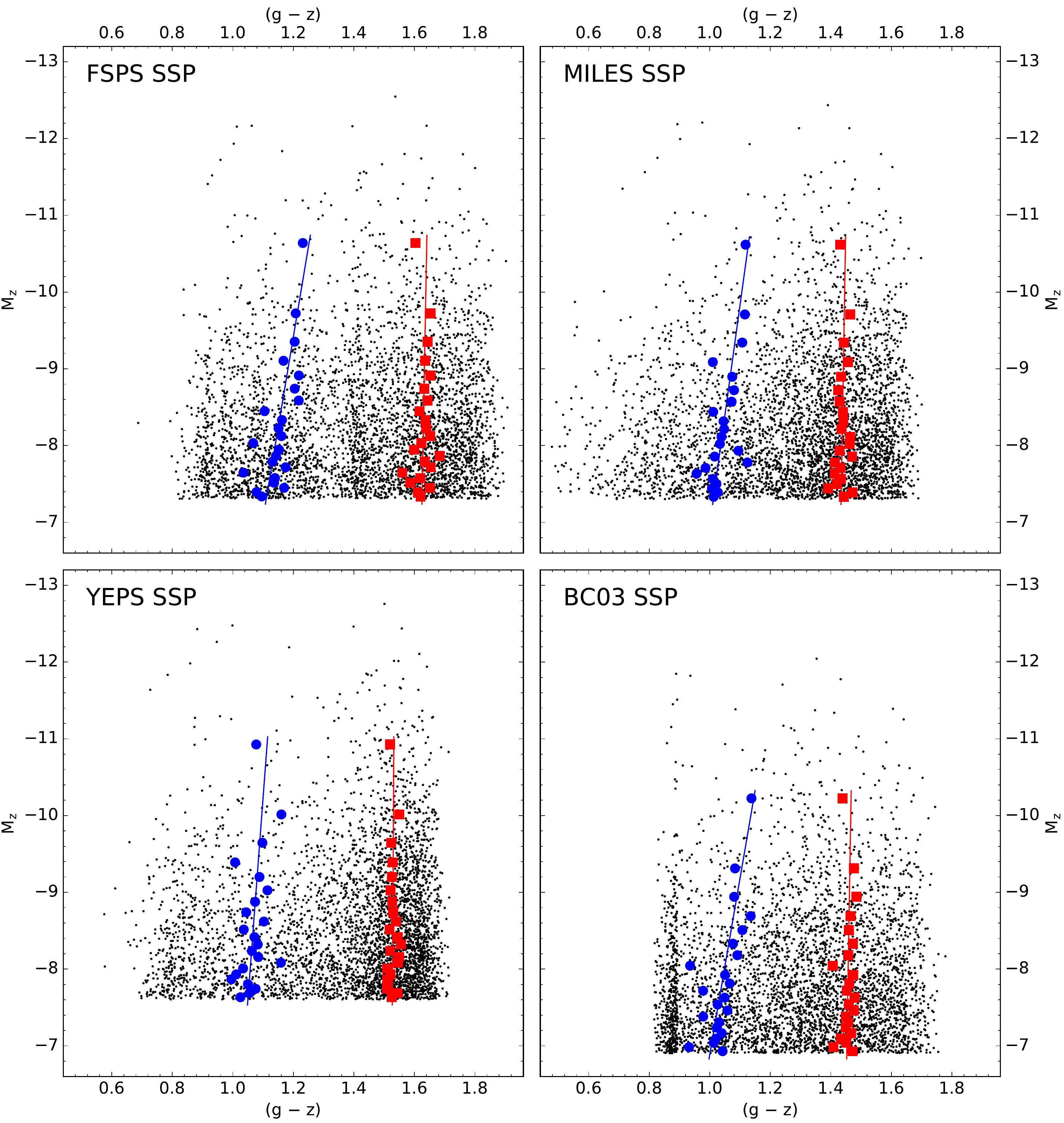}
\caption{Effects of different stellar population synthesis models on the blue tilt.
\emph{Top left}: Colours and absolute magnitudes calculated using the \citet{2009ApJ...699..486C} stellar population synthesis models.
\emph{Top right}: Colours and absolute magnitudes calculated using the \citet{2012MNRAS.424..157V} stellar population synthesis models.
\emph{Bottom left}: Colours and absolute magnitudes calculated using the \citet{2013ApJS..204....3C} stellar population synthesis models.
\emph{Bottom right}: Colours and absolute magnitudes calculated using the \citet{2003MNRAS.344.1000B} stellar population synthesis models.
While different stellar population models give different colour-magnitude relation slopes, the blue tilt is qualitatively the same as for the empirical colour-metallicity relation used in this work.
The range of colours and absolute magnitudes are the same in each panel and as in Figures \ref{fig:cmr_tests_metal} and \ref{fig:cmr_tests_cmr}.}
\label{fig:cmr_tests_ssp}
\end{center}
\end{figure*}

\begin{figure*}
\begin{center}
\includegraphics[width=504pt]{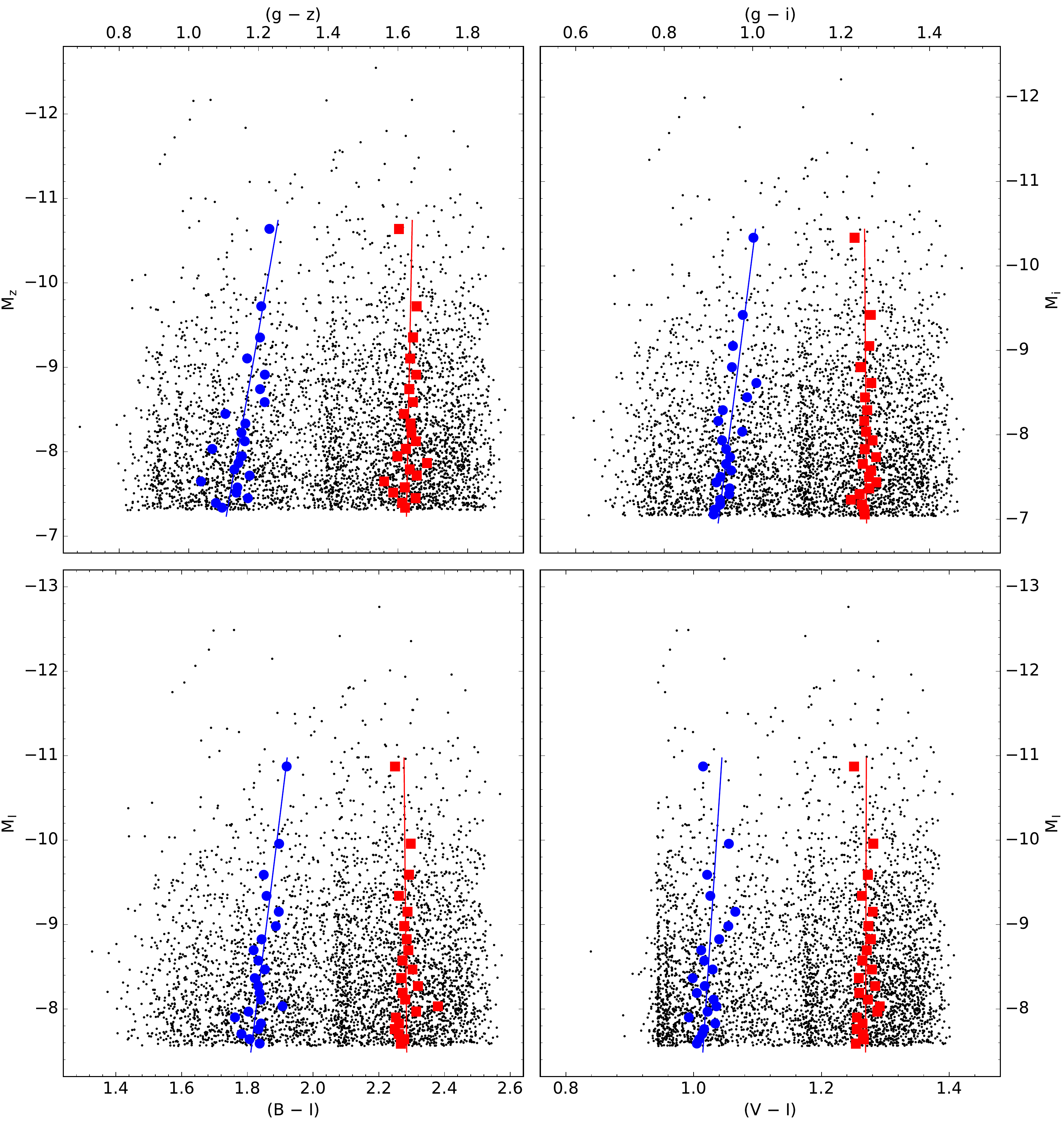}
\caption{Effects of different colours on the blue tilt.
\emph{Top left}: $(g - z)$ and $M_{z}$ calculated using the \citet{2009ApJ...699..486C} stellar population synthesis models.
\emph{Top right}: $(g - i)$ and $M_{i}$ calculated using the \citet{2009ApJ...699..486C} models.
\emph{Bottom left}: $(B - I)$ and $M_{I}$ calculated using the \citet{2009ApJ...699..486C} models.
\emph{Bottom right}: $(V - I)$ and $M_{I}$ calculated using the \citet{2009ApJ...699..486C} models.
While we see broadly similar blue tilts in $(g - z)$, $(g - i)$ and $(B - I)$, the blue tilt in $(V - I)$ is weaker due to the lack of metallicity sensitivity of this colour at low metallicity in the \citet{2009ApJ...699..486C} models.}
\label{fig:cmr_tests_colour}
\end{center}
\end{figure*}

Thirdly, in Figure~\ref{fig:cmr_tests_ssp} we use linear interpolations of four different single stellar population synthesis models \citep{2009ApJ...699..486C, 2012MNRAS.424..157V, 2013ApJS..204....3C, 2003MNRAS.344.1000B} to calculate colours and luminosities as functions of age, metallicity and mass. 
For each colour-metallicity relation or stellar population model, while the numerical value of the slope blue tilt changes, the qualitative properties of the blue tilt remain.
Other than for the \citet{2013ApJS..204....3C} models, the differences between the mass-metallicity relations calculated from the colour-metallicity relations and the mass-metallicity relations directly measured is smaller than the uncertainty in the colour-magnitude slopes.
Lastly, we investigate in Figure~\ref{fig:cmr_tests_colour} how the choice of colour affects the blue tilt, by repeating our analysis using the \citet{2009ApJ...699..486C} models for $(g - z)$, $(g - i)$, $(B - I)$ and $(V - I)$.
While the $(g - z)$, $(g - i)$ and $(B - I)$ colours give qualitatively similar blue tilts and mass-metallicity relations, the $(V - I)$ blue tilt is much weaker due to the poor sensitivity of this colour to metallicity.
We give the slopes of all our fits in Table~\ref{tab:slopes_cmr}.
We note that the colour distributions predicted by the \citet{2006ApJ...639...95P} and \citet{2012MNRAS.426.1475U} empirical colour-metallicity relations and all the stellar population models, save the \citet{2012MNRAS.424..157V} model, show peaks due to rapid changes in slope of their colour-metallicity relations at some metallicities.

\begin{table*}
\caption{Colour-magnitude relations for different transformations to photometry}
\label{tab:slopes_cmr}
\begin{tabular}{lcccccc}
\hline
Transformation & $d(g - z)_{b}/dM_{z}$ & $\gamma_{b}$ & $p_{b}$ & $d(g - z)_{r}/dM_{z}$ & $\gamma_{r}$ & $p_{r}$ \\
(1) & (2) & (3) & (4) & (5) & (6) & (7) \\ \hline
Default \citep{2012MNRAS.426.1475U} CMR  & $-0.0342_{-0.0130}^{+0.0122}$ & $0.219^{+0.083}_{-0.078}$ & 0.022 & $-0.0054_{-0.0039}^{+0.0039}$ & $\hphantom{-}0.035^{+0.025}_{-0.025}$ & 0.076 \\
{[Z/H]} $-\ 0.3$ dex \citet{2012MNRAS.426.1475U} CMR & $-0.0374_{-0.0091}^{+0.0116}$ & $0.239_{-0.074}^{+0.058}$ & 0.008 & $-0.0042_{-0.0039}^{+0.0033}$ & $\hphantom{-}0.027_{-0.021}^{+0.025}$ & 0.102 \\
{[Z/H]} $+\ 0.3$ dex \citet{2012MNRAS.426.1475U} CMR & $-0.0377_{-0.0094}^{+0.0168}$ & $0.242_{-0.108}^{+0.060}$ & 0.012 & $-0.0075_{-0.0037}^{+0.0047}$ & $\hphantom{-}0.048_{-0.030}^{+0.024}$ & 0.035 \\
{[Fe/H]} \citet{2012MNRAS.426.1475U} CMR  & $-0.0437_{-0.0051}^{+0.0136}$ & $0.279_{-0.087}^{+0.033}$ & 0.003 & $-0.0030_{-0.0042}^{+0.0041}$ & $\hphantom{-}0.019_{-0.027}^{+0.027}$ & 0.248 \\
\citet{2006ApJ...639...95P} CMR & $-0.0519_{-0.0115}^{+0.0120}$ & $0.237_{-0.055}^{+0.053}$ & 0.002 & $-0.0070_{-0.0055}^{+0.0042}$ & $\hphantom{-}0.032_{-0.019}^{+0.025}$ & 0.063 \\
\citet{2010AJ....140.2101S} CMR & $-0.0489_{-0.0091}^{+0.0122}$ & $0.319_{-0.080}^{+0.059}$ & 0.003 & $-0.0065_{-0.0062}^{+0.0040}$ & $\hphantom{-}0.032_{-0.020}^{+0.030}$ & 0.069 \\
\citet{2014MNRAS.437.1734V} CMR & $-0.0426_{-0.0130}^{+0.0155}$ & $0.256_{-0.093}^{+0.078}$ & 0.006 & $-0.0128_{-0.0038}^{+0.0064}$ & $\hphantom{-}0.077_{-0.038}^{+0.023}$ & 0.008 \\
\citet{2009ApJ...699..486C} SSP & $-0.0424_{-0.0129}^{+0.0132}$ & $0.297_{-0.092}^{+0.090}$ & 0.002 & $-0.0047_{-0.0061}^{+0.0055}$ & $\hphantom{-}0.019_{-0.021}^{+0.024}$ & 0.182 \\
\citet{2012MNRAS.424..157V} SSP & $-0.0340_{-0.0102}^{+0.0156}$ & $0.206_{-0.094}^{+0.062}$ & 0.011 & $-0.0043_{-0.0041}^{+0.0043}$ & $\hphantom{-}0.024_{-0.024}^{+0.023}$ & 0.164 \\
\citet{2013ApJS..204....3C} SSP & $-0.0191_{-0.0119}^{+0.0100}$ & $0.104_{-0.054}^{+0.065}$ & 0.044 & $-0.0015_{-0.0035}^{+0.0032}$ & $\hphantom{-}0.011_{-0.024}^{+0.026}$ & 0.292 \\
\citet{2003MNRAS.344.1000B} SSP & $-0.0436_{-0.0112}^{+0.0140}$ & $0.287_{-0.092}^{+0.073}$ & 0.005 & $-0.0043_{-0.0090}^{+0.0049}$ & $\hphantom{-}0.019_{-0.022}^{+0.041}$ & 0.191 \\ 
\citet{2009ApJ...699..486C} SSP $(g - i)$ & $-0.0243_{-0.0078}^{+0.0070}$ & $0.156_{-0.045}^{+0.050}$ & 0.003 & $\hphantom{-}0.0013_{-0.0053}^{+0.0035}$ & $-0.008_{-0.023}^{+0.034}$ & 0.532 \\
\citet{2009ApJ...699..486C} SSP $(B - I)$ & $-0.0317_{-0.0111}^{+0.0097}$ & $0.203_{-0.062}^{+0.071}$ & 0.011 & $\hphantom{-}0.0024_{-0.0076}^{+0.0037}$ & $-0.015_{-0.024}^{+0.049}$ & 0.526 \\
\citet{2009ApJ...699..486C} SSP $(V - I)$ & $-0.0086_{-0.0115}^{+0.0041}$ & $0.055_{-0.026}^{+0.074}$ & 0.043 & $-0.0004_{-0.0033}^{+0.0018}$ & $\hphantom{-}0.002_{-0.012}^{+0.021}$ & 0.340 \\  \hline
\end{tabular}

\medskip
\emph{Notes}
Column (1): Transformation name. [Z/H]
Column (2): Slope of the colour-magnitude relation for the blue subpopulation.
Column (3): Power law index of the mass-metallicity relation calculated from the slope of the blue colour-magnitude relation.
Column (4): Bootstrap probability that the blue colour-magnitude slope is greater than zero.
Column (5): Slope of the colour-magnitude relation for the red subpopulation.
Column (6): Power law index of the mass-metallicity relation calculated from the slope of the red colour-magnitude relation.
Column (7): Bootstrap probability that the red colour-magnitude slope is greater than zero.
\end{table*}

\section{Blue tilts of individual galaxies}
Colour-magnitude diagrams showing the blue tilt individually for each of our galaxies with at least 150 GCs are presented in Figure \ref{fig:per_halo_blue_tilt}.
The properties of the galaxies are listed in Table \ref{tab:slopes}.

\begin{figure*}
\begin{center}
\includegraphics[width=504pt]{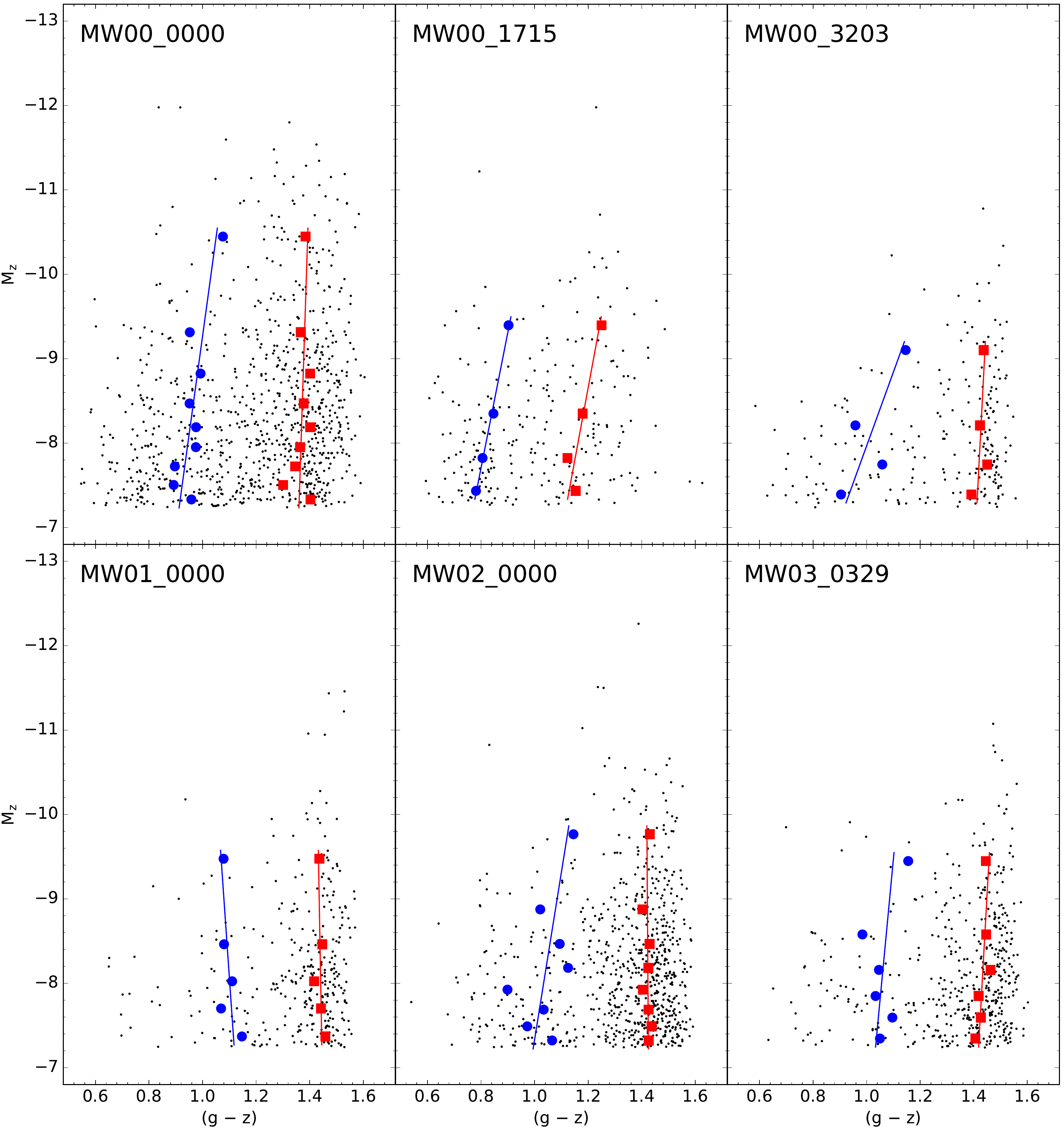}
\caption{Blue tilts of the individual galaxies with at least 150 old (age $> 8 \Gyr$) GCs more massive than $10^{5} \Msun$.
Each panel shows the colour-magnitude diagram for all GCs bound to that galaxy.
As in Figure \ref{fig:all_blue_tilt}, the blue circles and red squares show the means of two Gaussians fit to the colour distribution in equal sized bins of absolute magnitude while the blue and red lines are least squares fits to these mean colours as a function of absolute magnitude.
While most (but not all) galaxies show a significant blue tilt, few of the galaxies show significant evidence for a colour-magnitude relation of the red subpopulation.
The properties of the best-fitting lines are listed in Table~\ref{tab:slopes}.}
\label{fig:per_halo_blue_tilt}
\end{center}
\end{figure*}

\begin{figure*}
\begin{center}
\includegraphics[width=504pt]{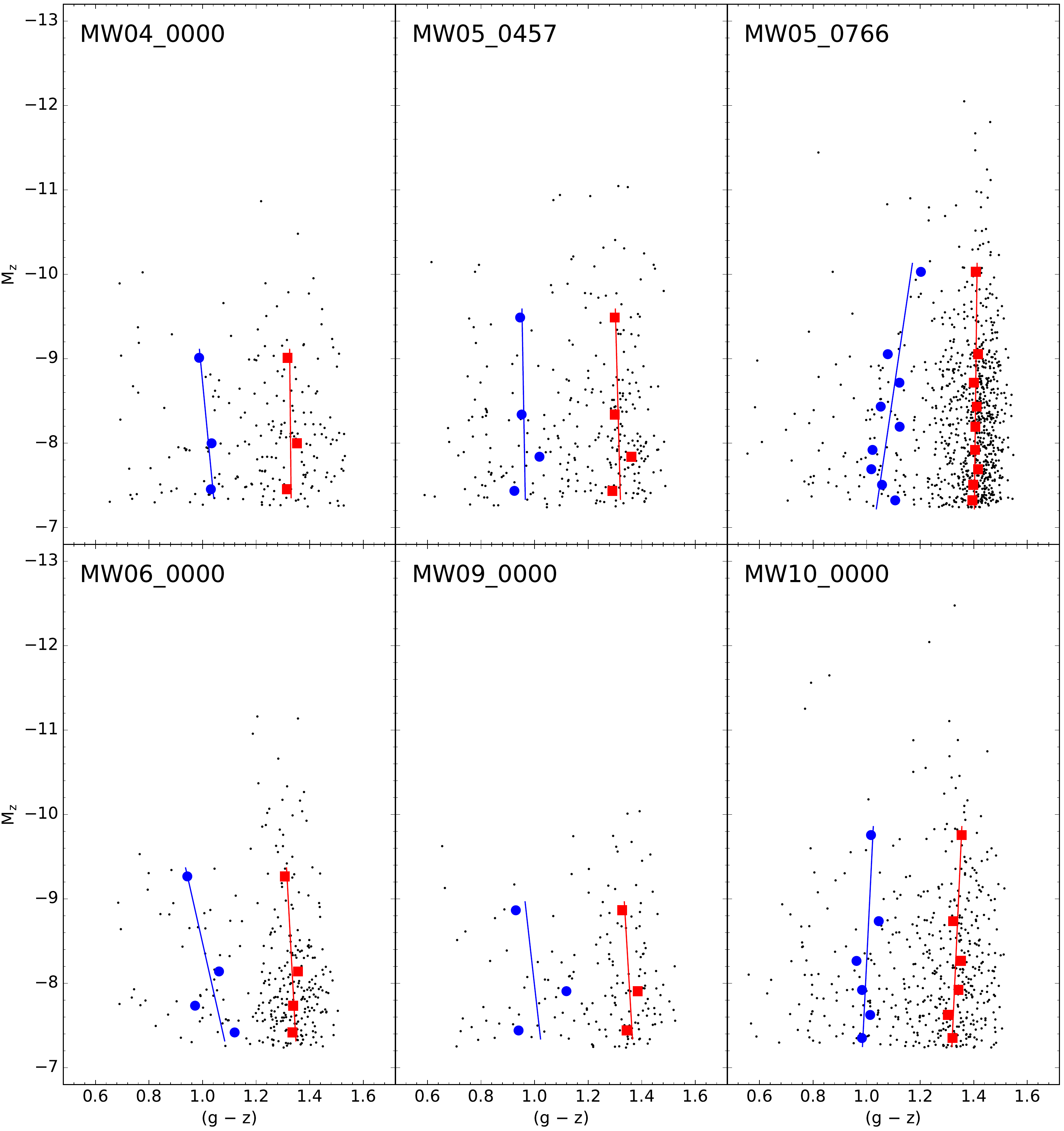}
\contcaption{Blue tilts of individual galaxies.}
\end{center}
\end{figure*}

\begin{figure*}
\begin{center}
\includegraphics[width=504pt]{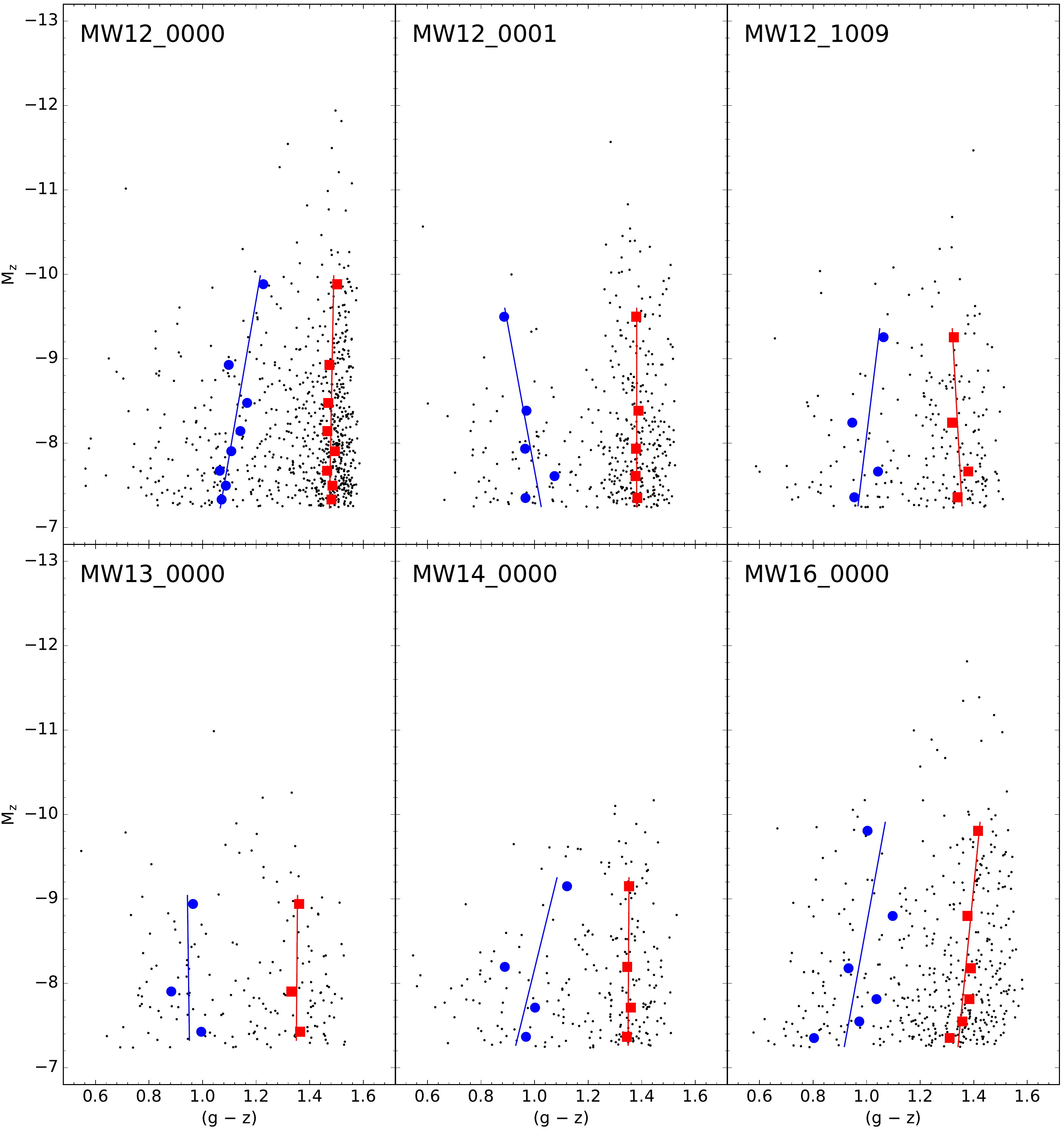}
\contcaption{Blue tilts of individual galaxies.}
\end{center}
\end{figure*}

\begin{figure*}
\begin{center}
\includegraphics[width=504pt]{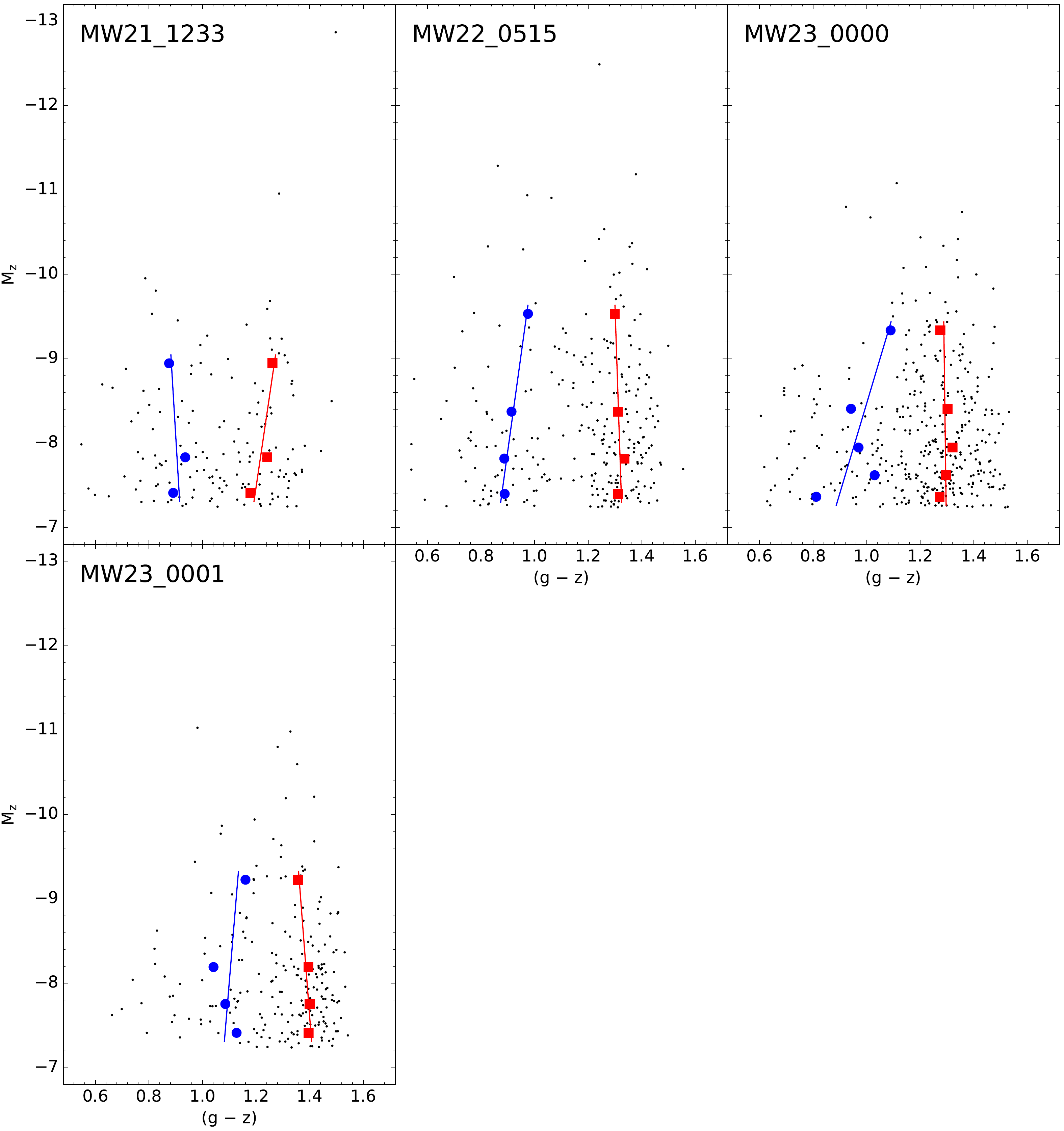}
\contcaption{Blue tilts of individual galaxies.}
\end{center}
\end{figure*}

\end{document}